\documentclass{aastex}
\usepackage{spr-astr-addons}
\usepackage{url}
\usepackage{graphics}
\usepackage{epsfig}
\usepackage{longtable}
\RequirePackage{color}
\begin{document}

\title{GRBs luminosity function synthesized from \textit{Swift/BAT}, \textit{Fermi/GBM} and \textit{Konus-Wind} data}

\author{H.~Zitouni}
\affil{PTEA Laboratory,  Faculty of Science, Dr Yahia Fares University, M\'{e}d\'{e}a, Algeria.
 \email{zitouni.hannachi@univ-medea.dz}}


 \and
\author{N.~Guessoum} \affil{Department of Physics, College of Arts \& Sciences, American University of Sharjah, UAE.}\email{nguessoum@aus.edu}
\and
\author{W.~J.~Azzam} \affil{Department of Physics, College of Science, University of Bahrain, Bahrain.} \email{wjazzam@sci.uob.bh}
 \and
{\author{Y.~Benturki}
\affil{PTEA Laboratory,  Faculty of Science, Dr Yahia Fares University, M\'{e}d\'{e}a, Algeria.}
\email{benturkiyassine@gmail.com}

\begin{abstract}

We study the luminosity function of long gamma-ray bursts (LGRBs) using the peak flux obtained from three LGRB samples with known redshifts: (a) a sample of 251  LGRBs from the \textit{Swift/BAT} satellite/instrument; (b) a sample of 37 LGRBs from the \textit{Fermi/GBM} telescope; (c) a sample of 152 GRBs from the \textit{Konus-Wind} instrument. For the \textit{Swift/BAT} and  \textit{Fermi/GBM}  samples, we use data available on the Swift Burst Analyser websites (\url{http://www.swift.ac.uk/burst_analyser};  \citep{Evans:2010}) and (\url{http://swift.gsfc.nasa.gov/archive/grb/table/})
and on the Fermi website (\url{https://heasarc.gsfc.nasa.gov/}\url{W3Browse /fermi /fermigbrst.html}; \citep{{Gruber_2014}, {von_Kienlin_2014}, {Bhat_2016}}) to calculate the luminosity at the peak of the flux by using a cut-off power-law spectrum (CPL).
For the \textit{Konus-Wind} sample, we use the Yonetoku correlation relationship \citep{Yonetoku:2010} to determine the isotropic luminosity from the energy at the peak of the flux measured in the source frame \citep{Minaev:2019}. With these three samples (totalling 439 GRBs), we use the Monte Carlo method to synthesize 10,000 ``artificial" GRBs similar to each real GRB by considering that each physical quantity observed obeys a normal distribution, where the tabulated value and uncertainty represent the mean and the $3\sigma$ error. The results obtained for the luminosity function from our data samples are consistent with those published in previous works.
\end{abstract}

\section{Introduction}
Gamma-ray bursts (GRBs) are extremely powerful explosions that involve either the death of massive stars or the merger of compact stellar objects. The equivalent isotropic energy of GRBs can exceed $ 10^{54}$ erg, and their nonthermal spectra peak between 10 and $10^4$ keV (e.g. \cite{Atteia_2017}). Their light curves display intense pulses that can last anywhere from less than a second to hundreds of seconds, they are often irregular and sometimes include multiple pulses. Although the radiation emitted by GRBs is believed to come out mainly through jets (with collisions inside), the precise formation mechanism behind these jets is not fully understood (e.g. \cite{Le_2017}).

Given the complex nature of their occurrence and unfolding, it is not surprising that GRBs are among the most studied topics and phenomena in astrophysics today, both theoretically and observationally. Many theoretical models have been proposed to try and explain this very spectacular phenomenon; but, unsurprisingly given the astounding amount of energy involved and the short time scales of these explosions, none of these models has been fully successful (\citep{Dainotti_2019}).

There are currently many observation missions, both from the ground and from space, that are involved in the study of GRBs in the different bands of the electromagnetic spectrum. Furthermore, a new observational tool has recently emerged, namely gravitational waves, which are produced by the merger of neutron stars or black holes. This new tool is already providing insight into the formation mechanisms of short, merger-type GRBs and has become an integral part of the multi-messenger approach to studying GRBs (\citealt{Abbott_2017}, \citealt{Drago_2018}, \citealt{Meszaros_2019}, and others).

The study of the luminosity function (LF), which is defined as the number of bursts per unit interval of luminosity, is part of the verification and validation of the theoretical models of the physical processes involved in the occurrence of GRBs. It allows us, among other things, to predict the production rates of GRBs and the probability of detecting gravitational waves associated with some of them. Furthermore, it can provide insights into the star-formation rate.

Previous research on luminosity functions has been done on different data samples, most of which contained a limited number of gamma-ray bursts \citep{{Schmidt:1999}, {Schmidt:2001}, {Sethi:2001}, {Yonetoku:04}, {Liang:2007}, {Virgili:2009},  {Butler:2010}, {Cao:2011}, {Salvaterra:2012}, {Wanderman:2015}, {Amaral:2016}, {Debdutta:2017}, {Kinugawa:2019}, {Lan:2019}}. Our present work uses three larger data samples from three satellites/instruments, namely Swift/BAT (251 GRBs), Konus/Wind (152 GRBs), and Fermi/BAT (37 GRBs), and then enlarges the data set with simulated (artificial) GRBs.

In Section 2, we describe how we select our samples. Sections 3 and 4 provide details of how we calculate the luminosities and the luminosity functions (for each sample), respectively. In Section 5, We present our results and then discuss them in Section 6, ending with concluding remarks and a mention of future research prospects in Section 7.

\section{Sample Selection}
\subsection{The Swift/BAT sample}
We use the \textit{Swift} GRBs data that is published on the satellite's official websites\footnote{\url{http://swift.gsfc.nasa.gov/archive/grb/table/},}$^,$\footnote{\url{http://www.swift.ac.uk/burst_analyser}}. The first one presents the observational results characterizing each detected GRB in several aspects: peak flux, fluence, duration, redshift, host galaxy, as well as data on afterglows (successive emission down the electromagnetic spectrum: X-ray, UV, visible, infrared, radio). The second website provides more details on the energy spectrum and the light curve (time profile) in different energy bands, with different time resolutions. The data are all provided with their uncertainties (error bars).

 As of 14/08/2020, \textit{Swift/BAT} had observed 379 GRBs with measured redshifts. We have eliminated five GRBs which redshifts were determined with poor precision: 060708, 090814A, 060116, 110726A, and 050904. Nine GRBs have no peak flux value in the published data: 050408, 130518A, 140614A, 150101B, 150821A, 151027B, 160117B, 170405A, and 170607A. Three bursts were removed due to the lack of one of the spectral parameters characterizing the cut-off power-law: 090516A, 140311A, and 071112C. With this filtering, we are left with 359 GRBs. Then, considering the \textit{Swift/BAT} energy band between 15 keV and 350 keV and following the same approach  as \cite{daigne:2006} and \cite{campisi:2010} in  taking  the lower limit of the peak flux as $P_{\gamma, lim}=0.2~~ph~cm^{-2}~s^{-1}$, we have a sample of 261 GRBs. However, considering the uncertainty on the edges of the energy band interval, it was possible to add three bursts: 120521C, 090715B, and 050416A. We get a sample of 265 GRBs (251 long GRBs and 15 short GRBs). Our final sample is then composed of 251 LGRBs, which full data (redshift, duration, spectral parameters) we present in Table (\ref{tabSwift}).
\subsection{The Fermi/GBM sample}
We have used the spectral data for Fermi/GBM GRBs given on the official website\footnote{\url{https://heasarc.gsfc.nasa.gov/W3Browse/fermi/fermigbrst.html}} \citep{{Gruber_2014}, {von_Kienlin_2014}, {Bhat_2016}}. Since we need GRBs with redshifts, we take the set given by \cite{Minaev:2019} and its references, as well as GCN (Gamma-ray Coordination Network) alerts. This gives us a total of 37 LGRBs, which we present in Table (\ref{tabFermi}).

\subsection{The Konus-Wind sample}

The \textit{Konus-Wind} satellite has observed 152 LGRBs \citep{{Svinkin:2016}, {Tsvetkova:2017},{Cano:2017}, {Minaev:2019},{Frederiks_2019}}. In this sample, we know the bursts' redshifts and energies at the peak $E_p$. We present this data in Table (\ref{tabKW}) to facilitate future research.

\subsection{Representativeness of the GRB sample (with known redshifts)}
Before we perform our calculations, we must ensure that the GRB samples (with known redshifts) that we will be using are representative of the whole population of long gamma-ray bursts (with or without known redshifts).

We first analyze the bursts in the Swift/ BAT sample. Our approach consists in comparing the distributions of the three physical quantities: the duration, $T_{90}^{Obs}$, the energy at the peak flux, $E_p^{Obs}$, and the flux at the peak. Taking into account the lower limit of the peak flux as $P_{\gamma, lim}=0.2~~ph~cm^{-2}~s^{-1}$) and the energy band of Swift/BAT (15-350 keV), the full sample, denoted by A (for All), is reduced to 870 LGRBs. The second sample, denoted by B (bursts of Known redshift z), consists of 251 GRBs (there are thus 619 LGRBs without known redshifts). For each of the above physical quantities, a comparison is made between its distribution for the full sample and that for the B sample (of bursts with known redshifts).

We present a visual comparison of the distributions by graphical representation in the figures (\ref{Distributions_Swift}: a, b and c).

\begin{figure}
       \centering
       \includegraphics[angle=0, width=0.45\textwidth]{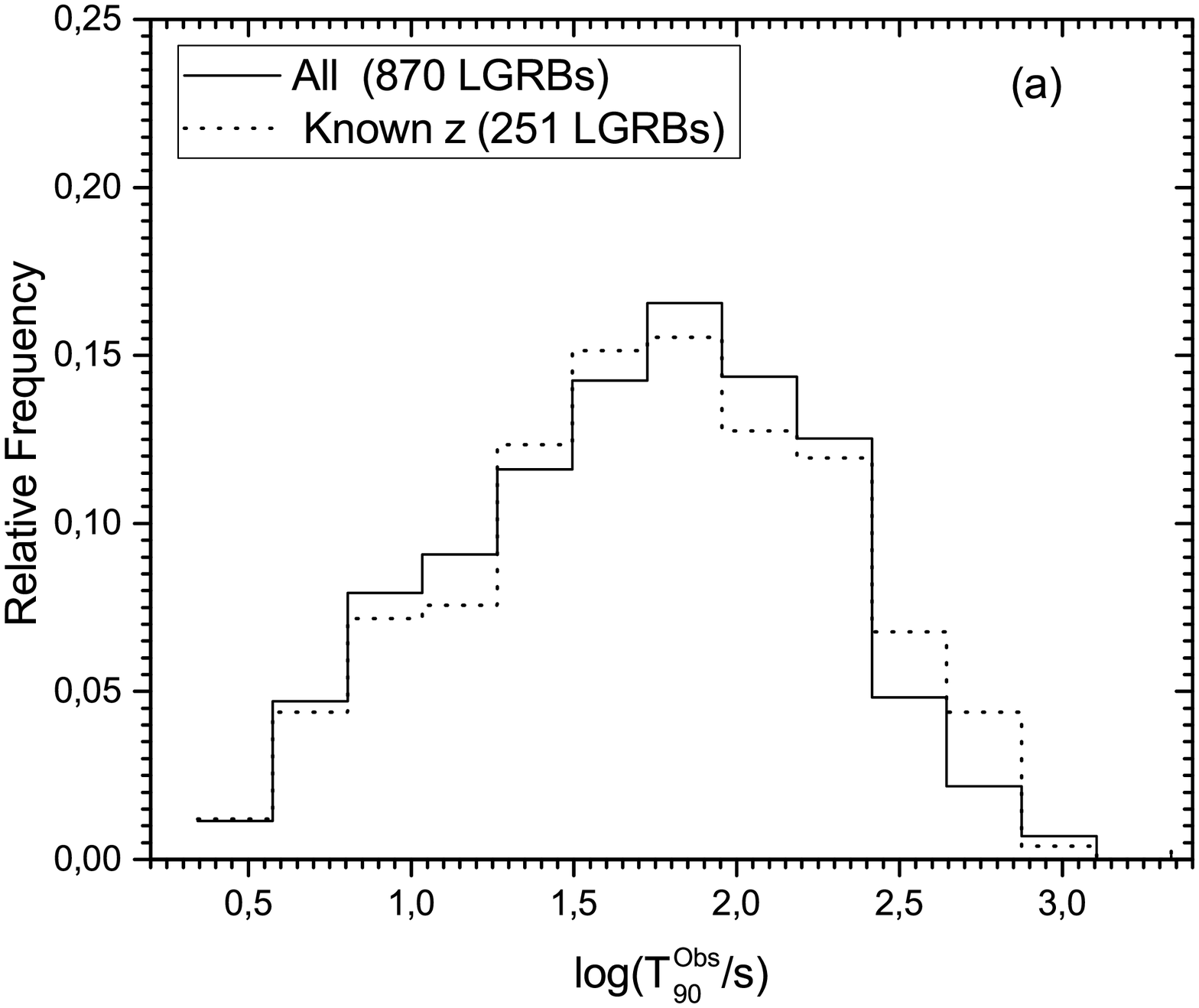}
       \includegraphics[angle=0, width=0.45\textwidth]{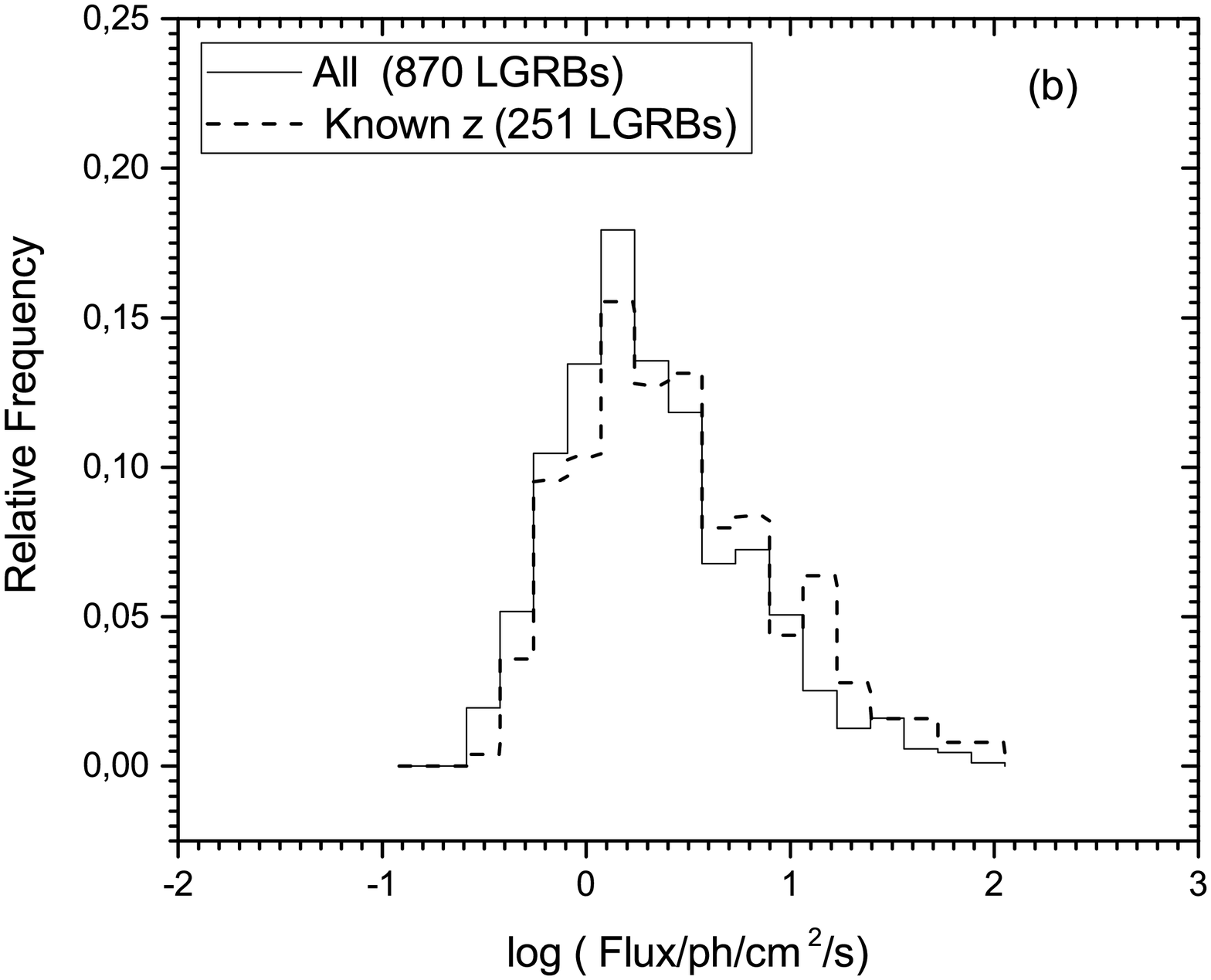}
       \includegraphics[angle=0, width=0.45\textwidth]{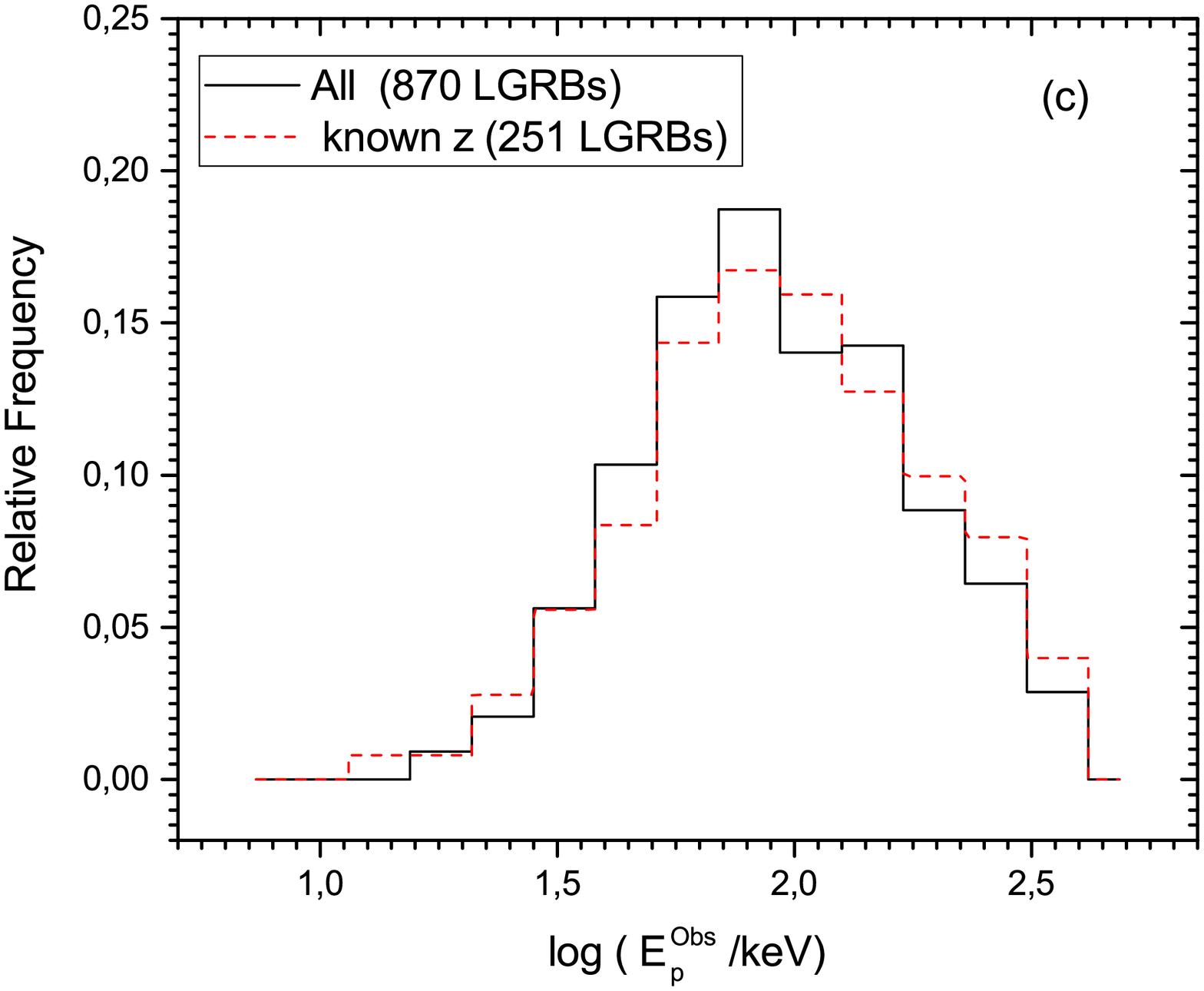}
        \caption[Distribution]{\emph{ (a): Duration distribution of  \textit{Swift/BAT} LGRBs; (b): Peak flux distribution of \textit{Swift/BAT} LGRBs; (c): $E_{peak}$ energy distribution of CPL for \textit{Swift/BAT} LGRBs.}}
        \label{Distributions_Swift}
\end{figure}

Our statistical analysis starts with a check of the normality of each of the six distributions (three for each sample). We adopted the Kolmogorov Smirnov test and applied it to each of the distributions, taking the logarithm of the physical quantities. The results obtained for each sample are presented in the two tables (\ref{normality_tab}).
We should note that the graphical representation was made on the normalized distributions, while the statistical tests were made on the actual ones.

\setcounter{table}{3}
\begin{table}[ht]
  \centering
  \footnotesize{
  \begin{tabular}{cccc}
    \hline
    &Statistic&p-value&Decision at 5\% level \\
    \hline
    \end{tabular}

  Sample of 870 LGRBs (A: All)
  \begin{tabular}{cccc}
   $\log{(T^{Obs}_{90})}$ &0.0445&0.063&can't reject normality\\
   $\log{(Flux)}$&0.0723&0.0002&reject normality\\
   $\log{(E_p^{Obs})}$&0.0322&0.328&can't reject normality\\
    \hline
   \end{tabular}
    Sample of 251 LGRBs (B: Known z)\\
    \begin{tabular}{cccc}
   $\log{(T^{Obs}_{90})}$ &0.041&0.864&can't reject normality\\
   $\log{(Flux)}$&0.077&0.098&can't reject normality\\
   $\log{(E_p^{Obs})}$&0.039&0.929&can't reject normality\\
    \hline
  \end{tabular}
  \caption{Normality tests for six distributions for two  Samples: A (All) composed of 870 LGRBs, and B (with know z) composed of 251 LGRBs.}
  \label{normality_tab}
  }
  \end{table}

  As the normality of the two quantities ($log (E_p^{Obs})$ and $log (T_{90}^{Obs})$ was not rejected at a 5\% significance level for the two samples, we proceeded to an F-test to check for the null hypothesis of equality of variances ($H_0: \sigma_{B}^2=\sigma_{A}^2$), and a t-test to check for the null hypothesis of equality of means $H_0: \mu_{B}=\mu_{A}$).The results are presented in the table (\ref{test_tab}).

  \begin{table}[ht]
  \centering
  \footnotesize{
  Hypothesis of equal variance\\
  $H_0: \sigma_{A}^2 = \sigma_{B}^2$
  \begin{tabular}{cccc}
    \hline
    &F-Statistic&p-value&Decision at 5\% level \\
    \hline
   $\log{(T^{Obs}_{90})}$ &0.916&0.377&can't reject $H_0$\\
   $\log{(E_p^{Obs})}$&0.869&0.156&can't reject $H_0$\\
    \hline
  \end{tabular}
  Hypothesis of equal means\\
$H_0: \mu_{A} = \mu_{B}$
  \begin{tabular}{cccc}
    \hline
    &t-Statistic&p-value&Decision at 5\% level \\
    \hline
   $\log{(T^{Obs}_{90})}$ &-1.223&0.221&can't reject $H_0$\\
   $\log{(E_p^{Obs})}$&-0.52&0.60&can't reject $H_0$\\
    \hline
   \hline
  \end{tabular}
  \caption{F-test for the hypothesis of the equality of variances and t-test for the hypothesis of the equality of means.   $F_{0.025,869,250}=1.2273$, $t_{0.25,1119}=1.962$.}
  \label{test_tab}
  }
  \end{table}
  We conclude that for the two quantities $log (E_p^{Obs})$ and $log (T_{90}^{Obs})$ the distributions given by the two samples are not significantly different at the 0.05 level.

  For the distribution of the logarithm of the flux, we performed a nonparametric Kolmogorov-Smirnov (KS) test, which rejected the null hypothesis (equality of the two distributions representing the two samples) at a 0.05 significance level, but not at a 0.005 level (p-value = 0.0059). On the other hand, a Mann-Whitney test concluded that the two distributions are significantly different at a level of 0.005 (p-value = $8.9~ 10^{-4}$).

  We should also note that in this analysis, we did not take the uncertainties on the measured quantities into account; perhaps if the uncertainties are considered, the distributions of flux could also have an acceptable non-significant difference.

  It can be concluded from this statistical analysis that the two samples A and B can be considered as drawn from the same population of LGRBs.

For our sample of Fermi/GBM bursts with known redshifts, due to their limited number (37 LGRBs), we found it useful to validate its representativeness by checking the Yonetoku relation on this sample, as this correlation is widely assumed to be true for all long bursts; it has been studied by a number of researchers and for several samples. We also tested it on our Swift/BAT sample (251 LGRBs), for which we found parameters (intercept and slope) very close to the original values published in \cite{Yonetoku:04}. Our results are presented graphically in figure \ref{Yonetoku_relation_12}(top). For the Fermi sample (37 long GRBs), the values we obtained for the intercept and slope parameters are within the error bars of those of the original relation. This is shown graphically in figure \ref{Yonetoku_relation_12}(bottom). For a clear comparison with the  ($\alpha$ and q) parameters of the relation \eqref{eq_YoRN}, we present the table below(\ref{table6}):

\begin{table}[ht]
  \centering
  \begin{tabular}{cccc}
    \hline
     &$\alpha$ & $q$  \\
    \hline
     Swift&$0.56^{+0.64}_{-0.29}$ &$1.95\pm0.13$ \\
     Fermi&$1.00^{+2.16}_{-0.68}$&$1.98^{+0.40}_{-0.20}$ \\
    Ref1 &$2.43^{+2.29}_{-1.76}$&$2.0\pm 0.2$\\
    Ref2 &$22.4^{+17.0}_{-9.7}$&$1.60\pm{0.08}$\\
    \hline
  \end{tabular}
  \caption{Yonetoku relation. Ref1: \cite{Yonetoku:04};
Ref2: \cite{Yonetoku:2010}}
\label{table6}
  \end{table}

\begin{figure}
       \centering
       \includegraphics[angle=0, width=0.45\textwidth]{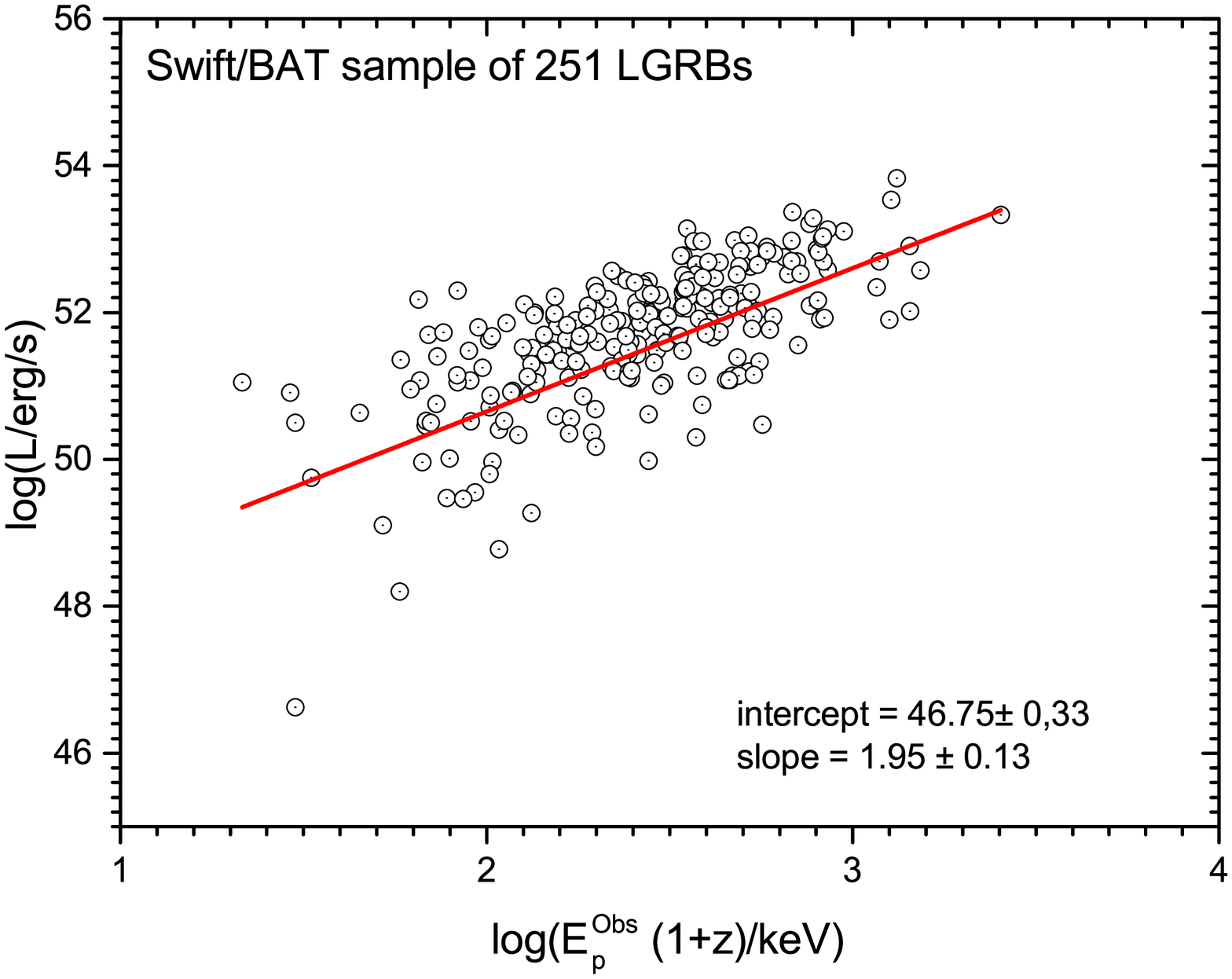}
         \includegraphics[angle=0, width=0.45\textwidth]{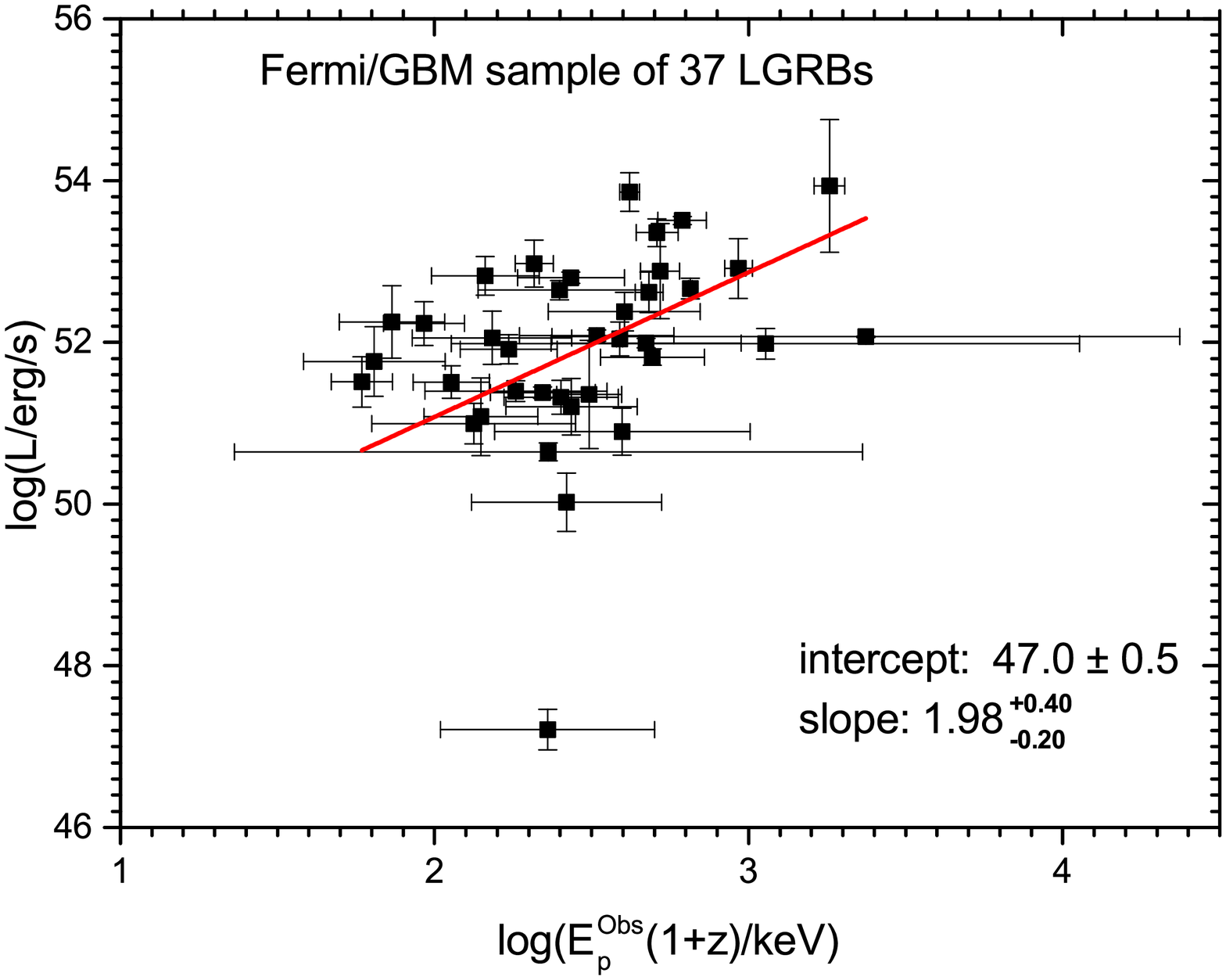}
        \caption[Distribution]{\emph{ Yonetoku relation: Luminosity versus intrinsic peak energy for
Swift/BAT (top) and Fermi/GBM (bottom) samples. In order not to clutter up the top figure, we have not shown the
error bars.}}
        \label{Yonetoku_relation_12}
\end{figure}

For the Konus-Wind sample, which consists of 152 LGRBs with known redshifts, we first check for the normality of the distributions of the duration and the observed energy at the peak flux, $E^{obs}_p$; the (normalized) distributions (of the logarithms of each quantity) are shown in the figures (\ref{Distributions_Konus_Wind} a and b). For this, we have performed a Kolmogorov Smirnov test on the data; the results we obtained are presented in the table (\ref{normality_tab_KW}).

\begin{figure}
       \centering
       \includegraphics[angle=0, width=0.45\textwidth]{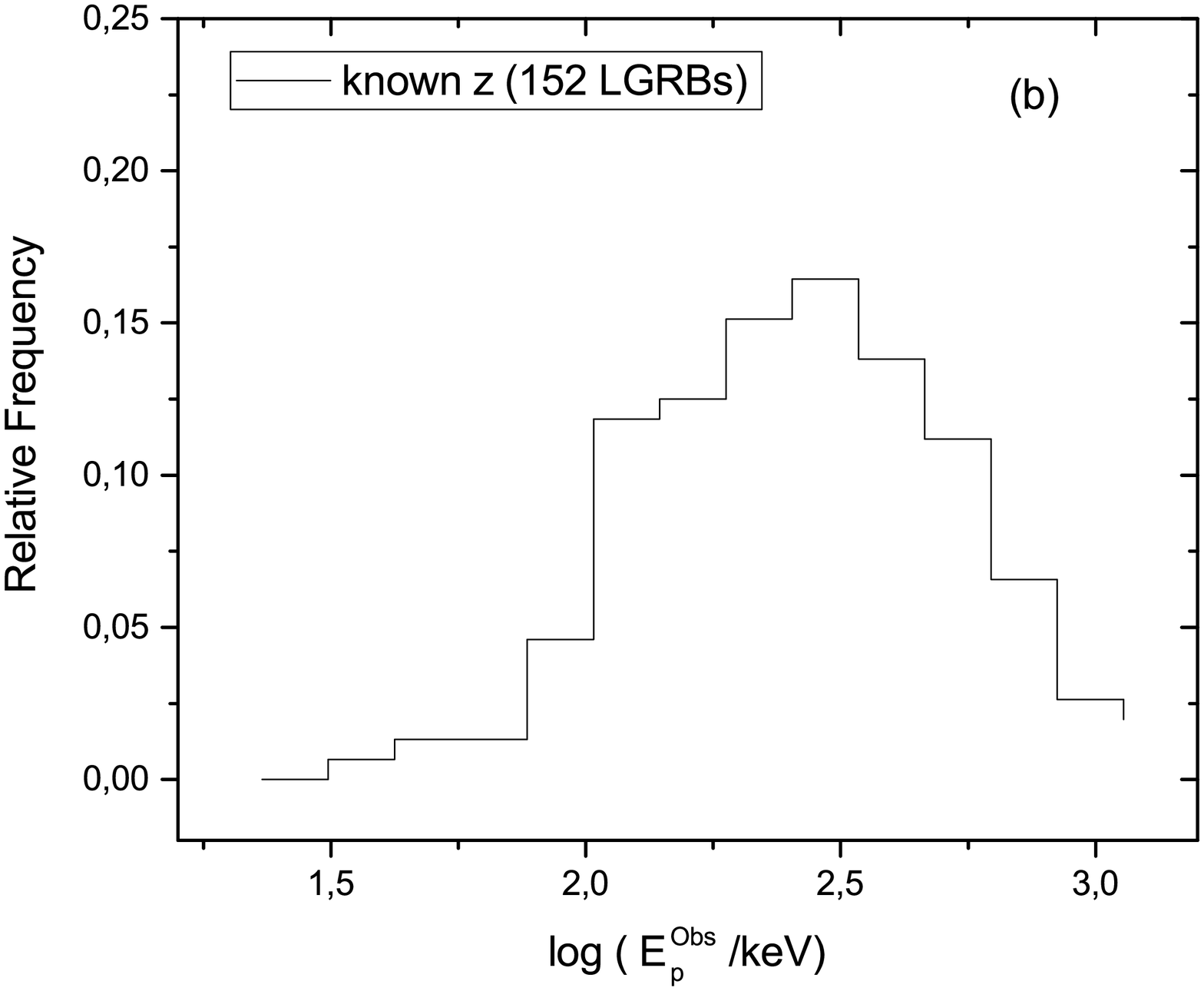}
       \includegraphics[angle=0, width=0.45\textwidth]{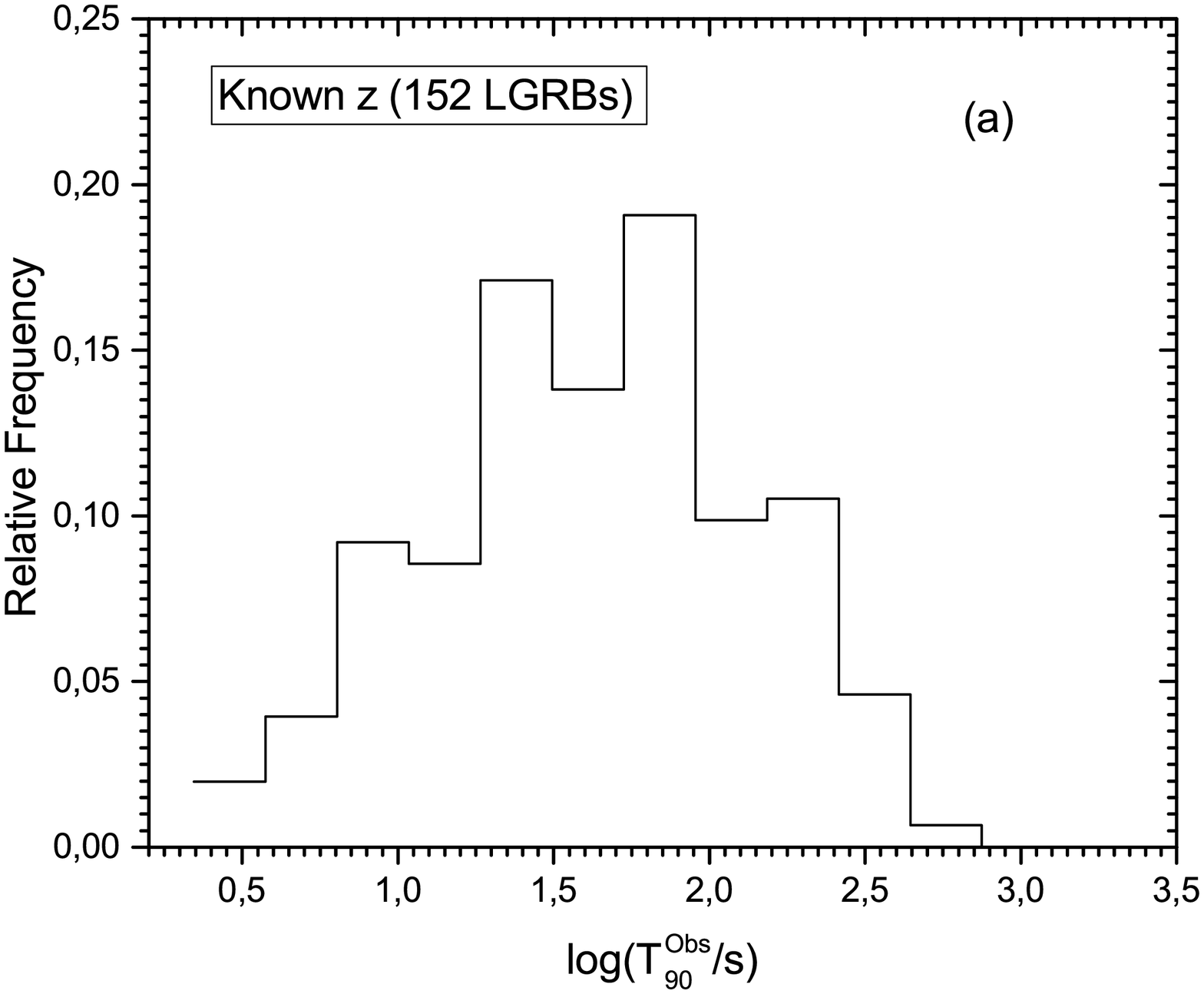}
        \caption[Distribution]{\emph{ (a): Duration distribution of  \textit{Konus-Wind} 152 LGRBs; (b): Peak flux distribution of \textit{Konus-Wind} 152 LGRBs.}}
        \label{Distributions_Konus_Wind}
\end{figure}

\begin{table}[ht]
  \centering
  \footnotesize{
  \begin{tabular}{cccc}
    \hline
    &Statistic&p-value&Decision at 5\% level \\
    \hline
   $\log{(T^{Obs}_{90})}$ &0.0406&0.99&can't reject normality\\
   $\log{(E_p^{Obs})}$&0.0327&0.99&can't reject normality\\
    \hline
   \end{tabular}
  \caption{Normality tests for two distributions for Konus-Wind  Sample with know z composed of 152 LGRBs.}
  \label{normality_tab_KW}
  }
  \end{table}

  Now to check for the representativeness of the Konus-Wind sample data, we use a comparison test of our sample against referenced values for the mean and variance as typically adopted for LGRBs. We use a t-test for the means and a $\chi^2$-test for the variance at a significance level of 0.05.
  The distribution of $log( E_p ^ {Obs}) $ of BATSE LGRBs (bright  and faint) studied by \cite{Nava2011} is characterized by ($\mu_0 = 2.29$, $\sigma_0^2 = 0.07 $ ). Taking these values as a reference and comparing with our Konus-Wind sample data, the t-test gives a p-value =0.012 for the means, and the $\chi^2$-test gives a p-value = 0.012 for the variance. In both tests the null hypothesis is  rejected at a significance level of 0.05 but is not for 0.01. In the same way, The distribution of $log( E_p ^ {Obs}) $ of the Fermi LGRBs (bright  and faint) studied by \cite{Nava2011} is characterized by ($\mu_0 = 2.31$, $\sigma_0^2 = 0.10 $ ). The t-test gives a p-value =0.09 for the means, and the $\chi^2$-test gives a p-value = 0.50 for the variance. In both tests the null hypothesis ($H_0: \mu = \mu_0$ and $\sigma^2 = \sigma_0^2$) is not rejected at a significance level of 0.05.

  The duration distribution was studied by  \cite{Tarnopolski_2015} for 1566 long Fermi bursts; he found a normal distribution for $log(T_{90}^{Obs})$ characterized by ($\mu_0 = 1.475\pm0.005 , \sigma_0^2 = 0.23\pm 0.04 $). In the same way as for $log(E_p^{Obs})$, a t-test on the means gave us a p-value = 0.23,  while a $\chi^2$-test of the variance gave a p-value = 0.048. The duration distribution was also studied by \cite{Shahmoradi_2013} considering 1366 long Fermi bursts; he found ($\mu_0 = 1.5, \sigma_0^2 = 0.25$). In the same manner, we find p-value = 0.53 in a t-test on the means and p-value = 0.225 in a  $\chi^2$-test on the variance. In both tests the null hypothesis ($H_0: \mu = \mu_0$ and $\sigma^2 = \sigma_0^2$) is not rejected at a significance level of 0.05, considering the error bars.


Finally, the Konus-Wind sample's representativeness could not be tested for the photon  flux, as this parameter is not given   among the satellite/detector's data for all the LGRBs of our KW sample ; nor could we utilize the Yonetoku relation, since we use it to derive the luminosity (again because the Yonetoku relation is widely assumed to be true for all long bursts).

From the above statistical tests, we conclude that our samples of LGRBs with known redshifts are mostly, if imperfectly, representative of the general populations of LGRBs (with or without known redshifts) with their physical parameters (duration, energy at the peak of the flux, flux at the peak) as recorded by the Swift, Fermi, and Konus-Wind satellites.

\section{Luminosity calculation}
 For the Swift/BAT and Fermi/GBM samples, the luminosity at the peak of the flux can be calculated using
 the spectral data of each burst, characterized by a known redshift.   The 1-second photon flux at the peak can be found in the \textit{Swift/BAT} and data  \textit{Fermi/GBM}. This flux gives the total number of photons per unit area and per unit time, regardless of their individual energies. The \textit{Swift/BAT} energy spectrum is divided into four bands: 15-25 keV, 25-50 keV, 50-100 keV, 100-350 keV. The \textit{Fermi/GBM} covers the wider band 10-1000 keV.
In this work, we use a cut-off power-law \citep{band:92}, which is characterized by two spectral parameters: the observed peak energy, $E^{obs}_p$, and the spectral index alpha ($\alpha$):
\begin{equation}\label{eq1n0}
     N_E(E)=A E^{\alpha}~e^{-E/E_0},
   \end{equation}
where  $E_0 = E^{obs}_p/(2+\alpha)$.

The bolometric luminosity of the 1-second isotropic peak flux, denoted by $L_{p}$, is the maximum energy radiated per unit time in all space. It is calculated by integrating the $EN_E$ function in the energy band corresponding to the observed gamma radiation band in the source's frame, i.e. $E_1$ = 1 keV to $E_2$ = $10^4$ keV. And because of  cosmological effects, the corresponding observed energy band is: $ E_1/(1+z)$ to $E_2/(1+z)$.\\ Thus, the k-corrected $L$ is calculated via:

\begin{equation}\label{eq:LZ}
   L= 4~\pi~d_L^2~P_{\gamma}~k_c.
 \end{equation}
 Here $L$ is k-corrected with the method developed by
 \citep{bloom:01}, and $P_{\gamma}$ is the peak flux expressed as $ph~cm^{-2}~s^{-1}$.  The k-correction factor is defined by \citep{{{Yonetoku:04},rossi:08},{elliott:12}} as:
\begin{equation}
k_c= \frac{\int_{E_{1}/(1+z)}^{E_{2}/(1+z)}E
N(E)dE}{\int_{E_{min}}^{E_{max}} N(E) dE} \ .
 \end{equation}

The cosmological distance $d_L$ is given by the equation:
 \begin{equation}
  d_L=\frac{(1+z)c}{H_0}\int_0^z
  \frac{dz'}{\sqrt{\Omega_M(1+z')^3+\Omega_{\Lambda}}} \ ,
 \end{equation}
where we adopt the following cosmological parameters: $\Omega_M$ = 0.27, $\Omega_{\Lambda}$ = 0.73, and $H_0$ = 70 $\mathrm{km/s/Mpc}$ \citep{komatsu:09}.

These integrals are performed numerically using the time-resolved spectral parameters given by \textit{Swift/BAT} and \textit{Fermi/GBM}.

With this data, as explained above, we calculate the luminosity of each burst, taking into account the error on each quantity used.
The results of this calculation are presented in Tables \ref{tabSwift} and \ref{tabFermi}. The errors are evaluated using the Monte Carlo method.  The quantities needed in the calculation of the luminosity ($E_p$, $\alpha$ , flux) are assumed to follow a normal distribution. For each GRB, by making N random draws from these distributions, we obtain the two extremum values of the luminosity. The subtraction of these two values from the calculated value without uncertainty gives the errors.

For the Konus Wind sample, given the difficulty of obtaining the spectral parameters ($\alpha$, $P_{\gamma}$) for all the bursts, we preferred to use the correlation relation of \cite{Yonetoku:04}, which relates the luminosity to $E_p$. We used the data for z and $E_p $ given in \cite{Minaev:2019}. We also used the improved \cite{Yonetoku:04} correlation relation:
\begin{equation}\label{eq_Yonetoku}
   \frac{L}{10^{52}}=(2.34^{+2.29}_{-1.76})\times 10^{-5} \bigg{[}\frac{E_p(1+z)}{1~keV}\bigg{]}^{2.0\pm0.2} \ .
\end{equation}

In our Monte Carlo simulation, we draw values of $L$ using the following relation (which is the same as \eqref{eq_Yonetoku}):
\begin{equation}\label{eq_YoRN}
  \frac{L(j)}{10^{52}}=\alpha \times 10^{-5} \bigg{[}\frac{E_{p,j}(1+z)}{1~keV}\bigg{]}^{q},
\end{equation}
 where $L(j)$ and $E_{p,j}$ are the luminosity and the energy corresponding to the jth GRB in our data table.
To estimate the error on $L(j)$, we assume that $\alpha$ and $q$ are random quantities obeying a normal distribution, such that: $\alpha=2.34^{+2.29}_{-1.75}$ and $q = 2.0\pm0.2$. We also assume that the peak energy $E_p(j)$ is normally distributed. The results of this calculation are presented in Table \ref{tabKW}.

\section{Method for calculating the luminosity function}

The total number of gamma-ray bursts observed by a telescope  within an aperture solid angle $\Omega$ and during an effective period of of activity T (in years) is:
\begin{equation}\label{eq_rate}
  N_{T} = \frac{\Omega T}{4\pi}\int_{0}^{z}\frac{R_{GRB}(z)}{1+z}\frac{dV}{dz} dz \int_{L_{min}}^{L_{max}} \varphi(L) dL,
\end{equation}
where  $\varphi(L)$ is the luminosity function and $R_{GRB}(z)$  is the rate of occurrence of GRBs at redshift z expressed in $Mpc^{-3} yr^{-1}$. The factor (1+z) introduces the cosmological expansion factor and $\frac{dV}{dz}$ is the variation of the volume of the universe as a function of the redshift z, given by the equation :
\begin{equation}\label{eq_volume}
  \frac{dV}{dz}=\frac{c}{H_0}\frac{4\pi d_L^2}{(1+z)^2 \sqrt{\Omega_M(1+z)^3+\Omega_{\Lambda}}}.
\end{equation}
The GRB luminosity function is usually defined as the number of bursts per unit of luminosity \citep{{Sethi:2001},{daigne:2006},{Zitouni:08}, {Cao:2011},{Amaral:2016},{Debdutta:2017}}:
\begin{equation}\label{eq_LFg}
\varphi(L) = \frac{dN}{dL}.
\end{equation}
To predict the total number $N_T$ of gamma-ray bursts that will be observed by a telescope, we need to know the rate of occurrence of GRBs and their luminosity function $\varphi(L)$. The $R_{GRB}$ rate is very poorly known despite the fact that there are several attempts based on broad assumptions.


In the present work, we assume that the rate of long gamma-ray bursts is proportional to the star formation rate (SFR). Indeed, it is now accepted that long gamma-ray bursts are due to the collapse of massive stars, thus the burst rate should be proportional to the supernova rate through a constant k parameter      \citep{{Porciani:2001},{daigne:2006}}:
 \begin{equation}\label{eq_SN_GRB}
   R_{GRB} = k \times R_{SN},
 \end{equation}
where $R_{SN}$ is the Type II supernova comoving rate.
Moreover, these authors assume that the rate of supernovae is proportional to the rate of massive star formation:
\begin{equation}\label{eqSN_SFR}
  R_{SN}= 0.0122 \ M_{\odot}^{-1}\times R_{SFR},
\end{equation}
where $R_{SFR}$ is the star formation rate, which is determined by observations only up to a redshift $z \simeq 2 - 2.5$. Beyond this value, several hypotheses are made in the literature, from which we retain the three most common ones: $SFR_1$, where the star formation rate decreases beyond $z \geq2$; $SFR_2$, where the rate remains constant beyond $z \geq 2$; and $SFR_3$, where the rate continues to increase beyond $ z\geq 2$ \citep{daigne:2006}. These three hypotheses are expressed by the following relations \citep{Porciani:2001}:
    \begin{equation}
        SFR_1(z)= 0.323 \  h_{70} \  \frac{e^{3.4z}}{e^{3.8z}+45}\quad [M_{\odot}~yr^{-1}Mpc^{-3}]
        \end{equation}
    \begin{equation}
        SFR_2(z)= 0.160 \  h_{70} \  \frac{e^{3.4z}}{e^{3.4z}+22}\quad [M_{\odot}~yr^{-1}Mpc^{-3}]
        \end{equation}
    \begin{equation}
        SFR_3(z)= 0.215 \ h_{70} \  \frac{e^{3.05z-0.4}}{e^{2.93z}+15}\quad [M_{\odot}~yr^{-1}Mpc^{-3}]
        \end{equation}


In our work, we have found it useful to use an equivalent form of the luminosity function presented by \cite{Butler:2010}. This form can be considered  as a scale-free function \citep{Sethi:2001}. Moreover, we are interested in determining the parameters (detailed below) that characterize this luminosity function, so we will use a form of normalized luminosity function defined as follows:
\begin{equation}\label{eq_LFz}
\phi(L) = \frac{1}{N_T\times L} \frac{dN}{ d\log{L}}.
\end{equation}

To calculate the GRB luminosity function, we use the samples of GRBs (and their data) from the \textit{Swift/BAT}, the \textit{Fermi/GBM}, and the \textit{Konus/Wind } satellites/instruments. With the exception of the redshift, all the parameters used in our calculations are given with their respective uncertainties. We have assumed that each parameter value is a normally distributed quantity characterized by the density function:
\begin{equation}\label{eq_Normale}
  f(x)=\frac{1}{\sigma\sqrt{2\pi}} \exp{-\frac{(x-\mu)^2}{2\sigma^2}},
\end{equation}
where $\mu$ represents the mean value of the measured quantity, and $\sigma$ corresponds to its uncertainty, both given in Table \ref{LZ_table}. For example, for each GRB, the energy at the peak $E_p$ given in the table corresponds to $\mu$ and the $\delta E_p$ corresponds to $\sigma$.

\section{Results}
\subsection{Luminosity versus redshift}
We study the variation of the luminosity observed as a function of the redshift for the three samples. We represent these results graphically in Figures (\ref{fig1LZ_Swift}, \ref{fig2LZ_Fermi}, \ref{fig3LZ_KW}), representing each instrument and its GRBs. Several previous works have shown the existence of a relationship between the luminosity L and the redshift z in the form:
\begin{equation}\label{eqLZ}
   L = L_{0} (1+z)^{k},
\end{equation}
where $k$ is the slope (in log-log), and $L_0$ is the intercept at z near 0.
From the figures (\ref{fig1LZ_Swift}, \ref{fig2LZ_Fermi}, \ref{fig3LZ_KW}), one can notice that the data can indeed be fitted by a relation such as given in equation \eqref{eq:LZ}, although there is significant scatter, especially for \textit{Konus-Wind}.  The numerical results of this fit are presented in Table \ref{LZ_table}. Some values obtained from other references are also presented.

\begin{table}[ht] \label{LZ_table}
  \centering
  \begin{tabular}{cccc}
    \hline
     &$L_0$ & $k$  \\
     &$erg.s^{-1}$ & \\
    \hline
     Swift&$50.33\pm 0.05$ &$3.40\pm0.10$ \\
     Fermi&$50.4\pm0.3$ &$3.8\pm0.7$ \\
    Konus/Wind & $51.5\pm0.5$&$2.93\pm0.10$ \\
    Ref1 &$49.98\pm0.09$&$2.95\pm 0.19$\\
    Ref2 &&$2.1\pm{0.6}$\\
    Ref3&&$2.43^{+0.41}_{-0.38}$\\
    Ref4&&$2.5^{ }_{ }$\\
    Ref5&&$2.22^{+0.32 }_{-0.31 }$\\
    \hline
  \end{tabular}
  \caption{Luminosity variation with redshift z. Ref1: \cite{Deng_2016}; Ref2: \cite{Salvaterra:2012}; Ref3: \cite{Yu_2015}; Ref4: \cite{Pescalli_2016}; Ref5: \cite{Lan:2019}.}
  \end{table}

\begin{figure}
       \centering
       \includegraphics[angle=0, width=0.45\textwidth]{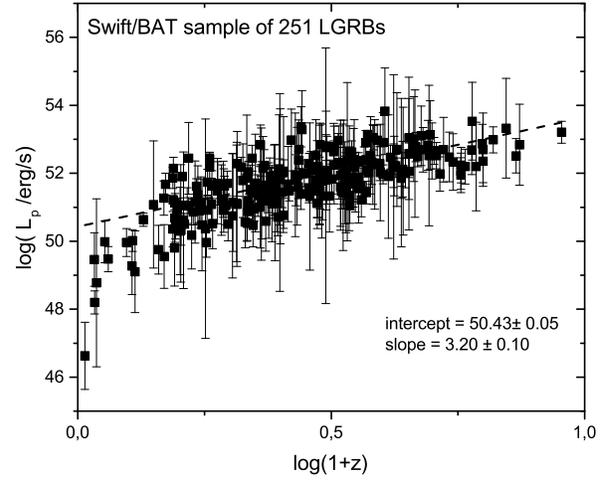}
        \caption[Luminosity versus redshift]{\emph{Luminosity  versus redshift for 251 \textit{Swift/BAT} LGRBs.}}
        \label{fig1LZ_Swift}
\end{figure}

\begin{figure}
       \centering
       \includegraphics[angle=0, width=0.45\textwidth]{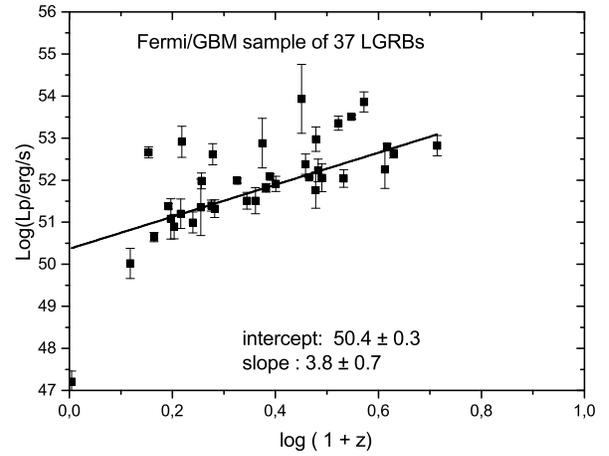}
        \caption[Luminosity versus redshift]{\emph{Luminosity  versus redshift for 37 \textit{Fermi/GBM} LGRBs.}}
        \label{fig2LZ_Fermi}
\end{figure}

\begin{figure}
       \centering
       \includegraphics[angle=0, width=0.45\textwidth]{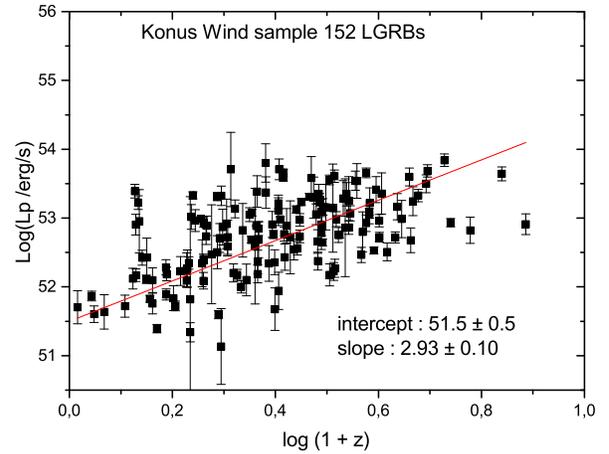}
        \caption[Luminosity versus redshift]{\emph{Luminosity  versus redshift for 152 \textit{Konus/Wind} LGRBs.}}
        \label{fig3LZ_KW}
\end{figure}

A luminosity evolution has been suggested by some authors, e.g. \citep{{Lloyd_Ronning_2002}, {Firmani_2004}, {Matsubayashi_2005}, {Kocevski_2006}, {Salvaterra_2009}, {Wanderman_2010}} with or without GRB rate evolution. Some of these authors used a sample of long bursts with known redshifts and found evidence for luminosity evolution. Other works find self-consistency without luminosity evolution. \cite{Butler:2010} conclude   that a rigorous treatment of the largest available Swift data set  allows for a firm conclusion in favor of rate evolution but not luminosity evolution. \cite{Deng_2016} say that with the large uncertainty on the $k$ index derived from one's simulation/analysis, one cannot convincingly argue for a robust evolution feature of the GRB luminosity.

\subsection{Luminosity function}

Based on the data from the three samples, we found that the luminosity function varies according to the broken power law form \citep{{Zitouni:08}, {Wanderman_2010}, {Butler:2010}, {Cao:2011},{Amaral:2016}}:
\begin{equation}\label{eq_LF}
 \phi(L) = \phi_0 \times
\left\{
  \begin{array}{ll}
    (\frac{L}{L_b})^{\nu_1}, & L \leq L_b \\
    (\frac{L}{L_b})^{\nu_2}, & L > L_b,
  \end{array}
\right.
\end{equation}
where $L_b$ is the luminosity at the break of the power law, and $\nu_1$ and $\nu_2$ are the exponents of each part of the function; $\phi_0$ corresponds to the value of the luminosity function at  the break $L=L_b$. $\phi(L)$ is expressed in $\frac{1}{erg.s^{-1}}$. Many authors have studied the evolution of the break luminosity $L_b$  and found that it varies with the redshift \citep{{Cao:2011},{Deng_2016},{Amaral:2016}}. The latter studies considered the evolution of the rate of formation of GRBs. For example, \cite{Amaral:2016} expressed the break luminosity as  $L_b= L_0 \times (1+z)^{\delta}$ and found that $\delta$ varies between $2.28$ and $4.09$ if we consider all models. For $L_0$, they found that it varies between $10^{50}~erg.s^{-1}$ and $10^{52}~erg.s^{-1}$.

For the luminosity function (LF) we present results with two cases/hypotheses: with evolution and without evolution of the LF. The ``unevolved" luminosity function is obtained in the same way that was done by \cite{Pescalli_2016}, taking $L_0 = L\times (1+z)^{-k}$. The parameter k is obtained from the table \ref{LZ_table} for each sample.

Using the likelihood method, we find the different parameters of the luminosity function for each sample and for each case ( evolution and non evolution of LF). The results are presented in Table \ref{LF_table} for the evolution case. For the Fermi sample, we could not get a fit with a broken power law..  These results are also represented graphically in the figures (\ref{fig2LF} a, b, c).

\begin{figure}
       \centering
       \includegraphics[angle=0, width=0.45\textwidth]{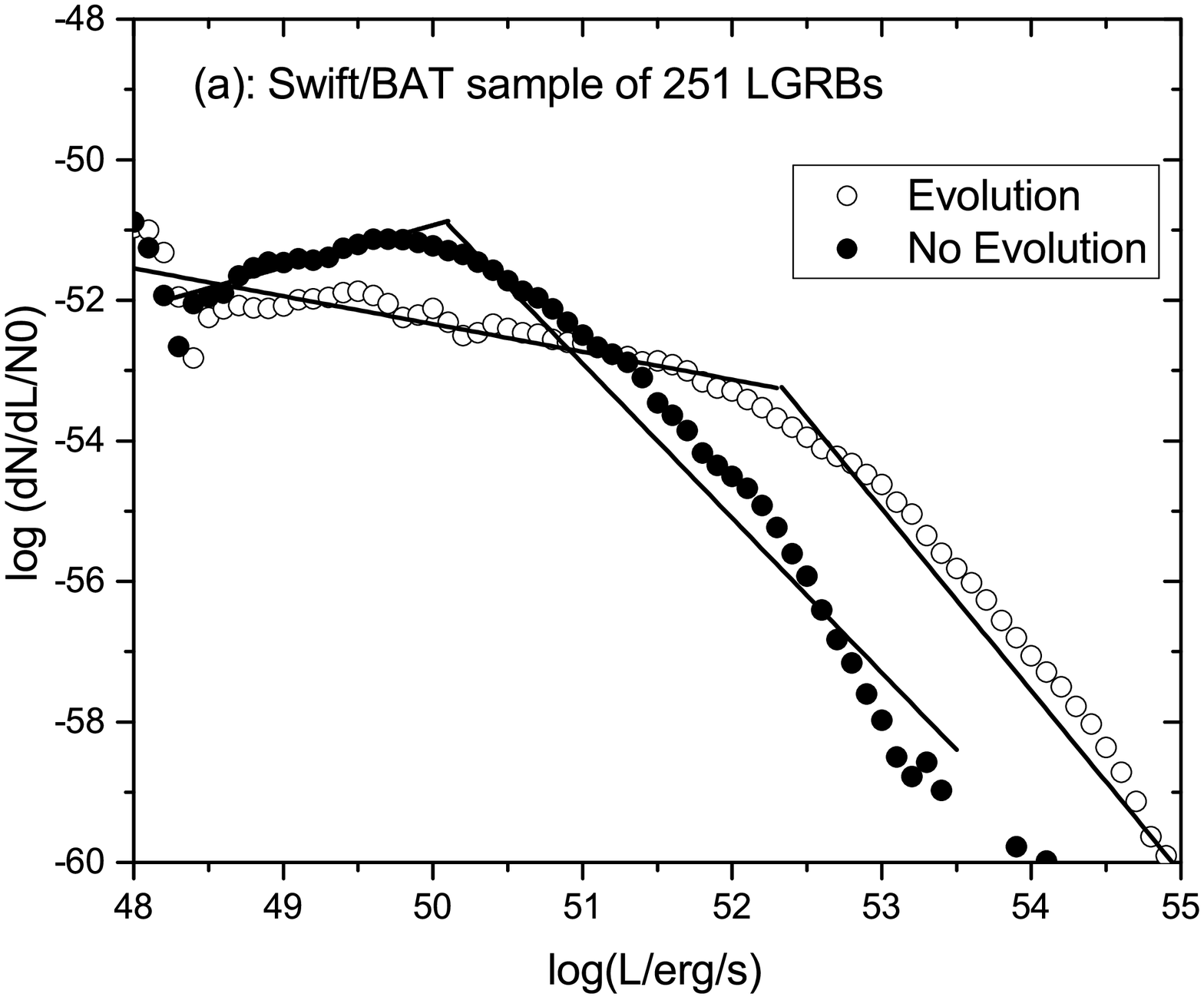}
       \includegraphics[angle=0, width=0.45\textwidth]{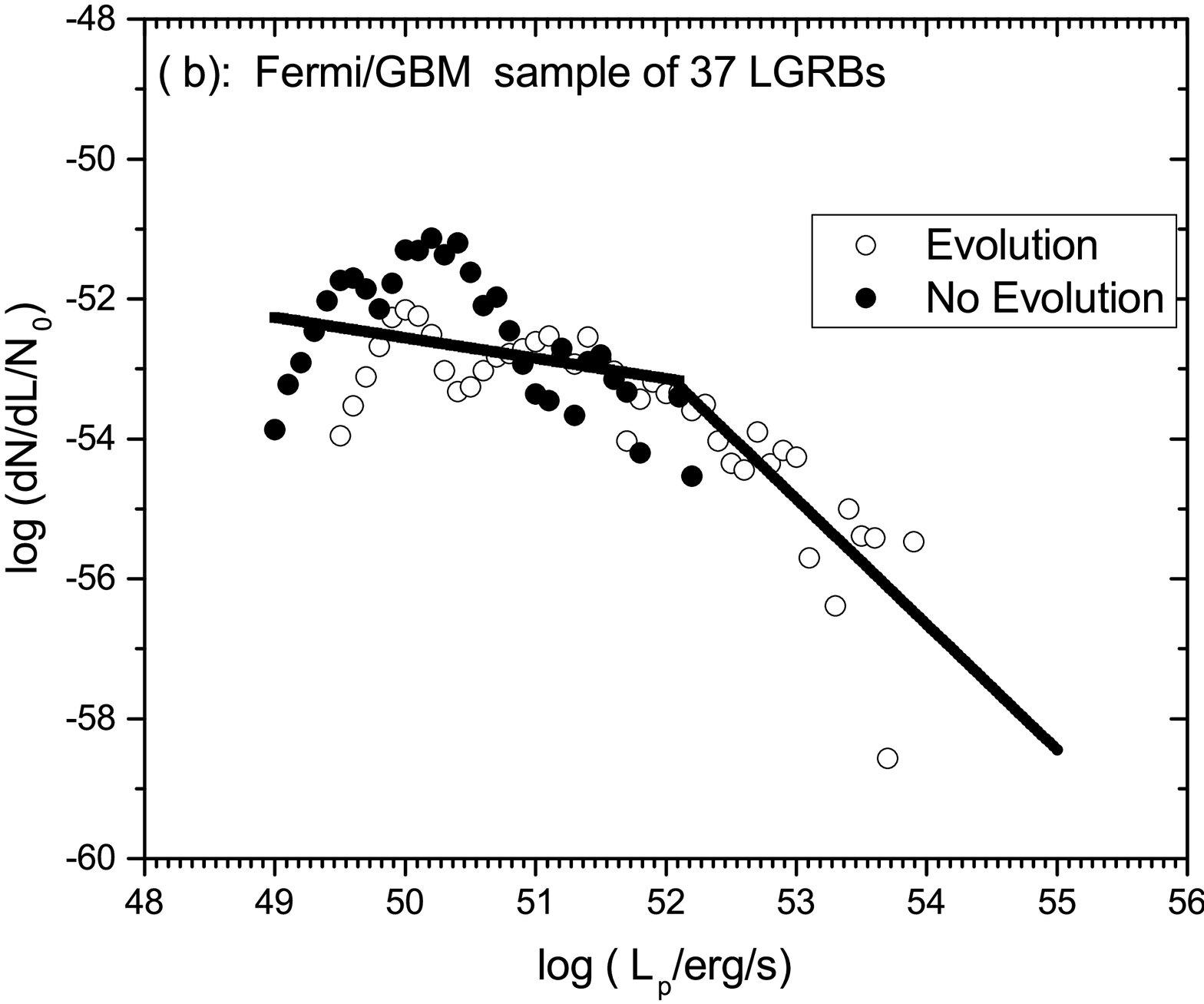}
       \includegraphics[angle=0, width=0.45\textwidth]{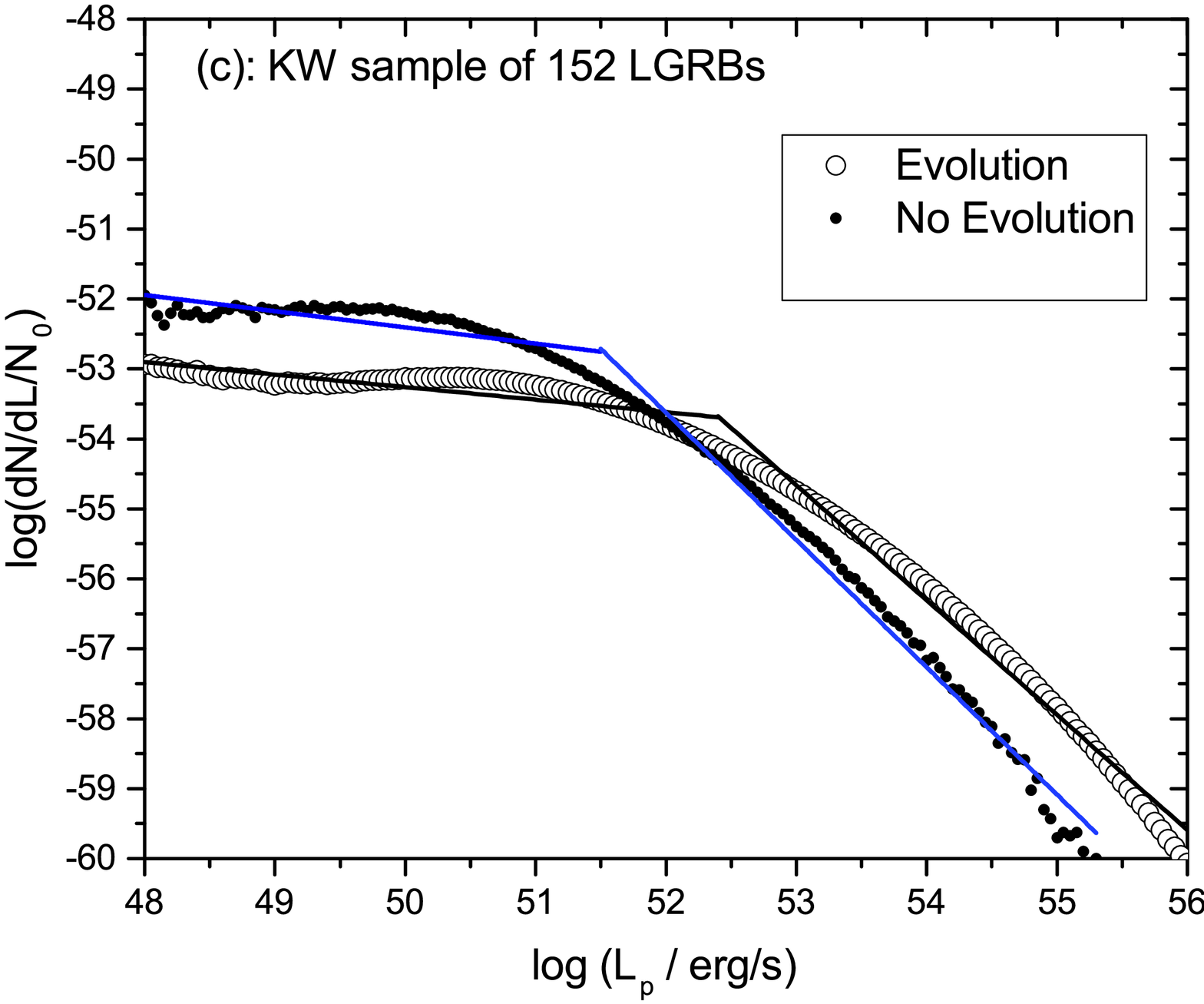}
        \caption[Luminosity Function]{Plots of the luminosity function obtained using: (a) 251 \textit{Swift/BAT} GRBs, giving $\log{(\phi_0)}=-52.66^{+0.09}_{-0.08}$; (b) 37\textit{Fermi/GBM} GRBs, giving  $\log{(\phi_0)}=-52.52^{+0.05}_{-0.07}$; (c) 152 \textit{Konus/Wind} GRBs, giving $\log{(\phi_0)}=-53.18^{+0.10}_{-0.8}$.}
        \label{fig2LF}
\end{figure}
\begin{table}[!ht]
  \centering
  \begin{tabular}{cccc}
    \hline
     &$\frac{\log{L_b}}{erg.s^{-1}}$ &~$\nu_1$& $\nu_2$    \\
     \hline
    No& evolution&&\\
     \hline
     &&&\\
    Swift&$50.1^{+0.1}_{-0.3}$ &$+0.60^{+0.08}_{-0.08}$&$-2.2^{+0.1}_{-0.2}$\\
    &&&\\
     Fermi&-&-&-\\
     &&&\\
     KW&$51.5^{+0.2}_{-0.20}$&$-0.23^{+0.02}_{-0.02}$&$-1.82^{+0.03}_{-0.03}$\\
     &&&\\
     Ref1&  $51.45^{+0.15}_{-0.15}$&$-1.32^{+0.21}_{-0.21}$&$-1.84^{+0.24}_{-0.24}$\\
     &&&\\
     Ref2&  $51.15^{ }_{ }$&$-0.14^{+0.02}_{-0.02}$&$-0.70^{+0.03}_{-0.03}$\\
     &&&\\
     Ref3& $52.22^{+0.08}_{-0.23}$&$-0.79^{+0.04}_{-0.04}$&$-1.91^{+2.04}_{-2.04}$\\
     &&&\\
     Ref4 & $52.17^{+0.18}_{-0.14}$&$-0.36^{+0.16}_{-0.16}$&$-1.28^{+0.09}_{-0.09}$\\
     &&&\\
    \hline
    With& evolution&&\\
    \hline
     &&&\\
    Swift&$52.3^{+0.2}_{-0.1}$ &$-0.40^{+0.03}_{-0.03}$&$-2.59^{+0.07}_{-0.08}$\\
    &&&\\
     Fermi&$52.1^{+0.1}_{-0.2}$&$-0.29^{+0.01}_{-0.01}$&$-1.79^{+0.04}_{-0.02}$\\
     &&&\\
     KW&$52.4^{+0.2}_{-0.2}$&$-0.18^{+0.01}_{-0.01}$&$-1.64^{+0.02}_{-0.02}$\\
     &&&\\
      Ref4 & $52.12^{+0.26}_{-0.26}$&$-0.72^{+0.05}_{-0.05}$&$-1.42^{+0.24}_{-0.24}$\\
     &&&\\
     Ref5&  $52.74^{+0.43}_{-0.43}$&$-0.22^{+0.31}_{-0.18}$&$-2.89^{+1.06}_{-2.05}$\\
     &&&\\
     Ref6 & $50.74^{+0.35}_{-0.42}$&$-0.74^{+1.36}_{-1.42}$&$-1.92^{+0.11}_{-0.14}$\\
     &&&\\
      Ref7 & $52.05^{+0.02}_{-0.02}$&$-0.20^{+0.1}_{-0.2}$&$-1.4^{+0.6}_{-0.3}$\\
      &&&\\
    \hline
  \end{tabular}
  \caption{Broken power law parameters of the luminosity function  (equation \eqref{eq_LF}). Ref1: \cite{Pescalli_2016}; Ref2: \cite{Yu_2015}; Ref3: \cite{Amaral:2016};   Ref4: \cite{Lan:2019}; Ref5: \cite{Butler:2010}; Ref6: \cite{Salvaterra:2012}; Ref7:  \cite{Wanderman_2010}.}
  \label{LF_table}
  \end{table}

\section{Discussion}
Let us now compare and contrast our work and results to those of other researchers.

Like our own, the investigation by \cite{Wanderman:2015} involved a joint analysis of GRBs from different satellites, with data from Swift, Fermi, and CGRO-BATSE, aiming to determine the luminosity function and the formation rate of GRBs. However, unlike our study, their investigation was restricted to short GRBs, and more specifically to non-collapsar short bursts, having found that their short-GRBs sample was ''contaminated'' by collapsars, which affected their data sample. With those restrictions, they obtained a luminosity function that was best fit with a broken power law with a break at $L_b = 2 \times 10^{52} erg.s^{-1}$, which is consistent with our results.

The investigation by \cite{Pescalli_2016} is most relevant to us because, like our study, theirs examines the luminosity function of GRBs and briefly explores whether GRBs undergo luminosity evolution. However, these authors restrict their data sample to long Swift GRBs, which reduces their sample to only 99 bursts compared to the 439 bursts in ours. Furthermore, our data is augmented by thousands of simulated ''artificial'' bursts for simulation purposes. Despite these differences, \cite{Pescalli_2016} find that their data can be best fit with a broken power law with $\log{(L_b)} = 51.45$, which is consistent with what we obtain. However, although their high luminosity index $\nu_2 (= -1.84)$ is consistent with our value, their low luminosity index $\nu_1 (= -1.32)$ is not, indicating that the low-end of their luminosity function is steeper than ours. The difference could be due to their data sample being restricted to long GRBs taken from Swift only, whereas our sample consists of data combined three sub-samples. We should note that a close examination of our results reveals that among our three sub-samples, the $\nu_1$ value for the Swift sub-sample was significantly the steepest among the three, which is at least qualitatively in agreement with \cite{Pescalli_2016}, and it might be hinting that the lower-end of the luminosity function is steeper for the Swift GRBs in particular. Moreover, like us, \cite{Pescalli_2016} found that there was a correlation between the luminosity and the redshift but concluded that this evolution should not be interpreted as being purely due to luminosity evolution. We agree with them, as the correlation with the redshift could be due to density evolution as well, and like \cite{Pescalli_2016} we admit that this issue needs further work, beyond the scope of our current study, because simulating both the luminosity and density evolutions and extricating the impact of each on the luminosity function and finding the extent and shape of its evolution is a separate and full-fledged topic. This is also the conclusion reached by \cite{Lan:2019} who analyzed the same data sample used by \cite{Pescalli_2016} to investigate the GRB luminosity function and utilize it to determine the star-formation rate.

The investigation by \cite{Amaral:2016} takes a distinct approach from ours. These authors explore the pulse luminosity function for Swift GRBs. They use a sample of 118 bursts to generate a total of 607 pulses and then compute the pulse luminosity function by employing three different GRB formation rate models. Despite this markedly different approach that includes contributions from all fitted pulses and not just from the brightest ones, the best-fit parameters for their luminosity function, $L_b = 1.5 \times 10^{51} erg.s^{-1}$, $\nu_1 = -0.70$, and $\nu_2 = -1.69$, are consistent with our results, although once again their low-end luminosity function is somewhat steeper than what we obtain.

\section{Conclusion}
The investigation of the GRB luminosity function is an important issue in GRB research because it gives insight into the physics involved in the production mechanism of GRBs. Not only does it shed light on the range of luminosities that GRBs can have, it also reveals whether there is a ''scale'' for GRB luminosities. The fact that all three sub-samples that we explored had a broken power-law luminosity function, with break luminosities consistent with one another, strongly indicates that the physics behind the formation of GRBs is not ''scaleless'', a result which can be useful when probing other physical aspects of GRBs.

In this study, we performed separate analyses of Swift/BAT, Fermi/GBM, and Konus/Wind data for all GRBs with known redshifts, in the aim of determining their luminosity functions. In doing that, we left out the formation rate of the GRBs from our considerations. We obtained the best parameters for the luminosity functions using the likelihood method. We found that they can be well represented by a broken power law -the so-called broken luminosity function.

The luminosity function expressed in log-log is well fit with a broken power-law with a break $L_b$ in the luminosity, which varies from $1.2 \times 10^{52} ~erg.s^{-1}$ to $ 6.3 \times 10^{52} erg.s^{-1}$. The exponent-index of the low luminosity part, $\nu_1$, varies between -0.23 and -0.47 and that of the high luminosity part, $\nu_2$, varies between -1.79 and -2.68. These power law indices are consistent with those found in previous studies (values and references are given in Table \ref{LF_table}).

In the future, we plan to look at both the density and luminosity evolution of GRBs, which is something that we only brushed briefly in this study. Such an investigation is important because it will cast light on several critical issues such as the star-formation rate, the degree to which GRBs can be utilized as ''standard candle'', and the evolution of the luminosity function itself.


\begin{acknowledgments}

 The authors gratefully acknowledge the use of the online \textit{Swift/BAT} table compiled by  Taka Sakamoto and  Scott D. Barthelmy  and \cite{Evans:2010}. The authors also aknowledge the use of the online Fermi/GMB table compiled by \cite{von_Kienlin_2014}, \cite{Gruber_2014} and \cite{Bhat_2016}. We thank the General Directorate of Scientific Research and Technological Development (DGRSDT), Algiers, Algeria for the financial support.
\end{acknowledgments}
\bibliographystyle{spr-mp-nameyear-cnd}
\bibliography{ref_Avril_2020}
\appendix
\setcounter{table}{0}
\scriptsize {
\begin{longtable}{|c|c|c|c|c|c|c|}
\caption{Data for 251 \textit{Swift/BAT}  GRBs.} \label{tabSwift}\\
  \hline
  GRB  & z & $T_{90}$ &  $E_p^{obs}$ & $\alpha$ & flux & $\log{\frac{L}{erg/s}}$ \\
   &   & sec & keV & &$ph/cm^2/s$ & \\
\hline
\endfirsthead
 \multicolumn{7}{c}%
{{\bfseries \tablename\ \thetable{} -- continued from previous page}} \\
\hline  GRB  & z & $T_{90}$ &  $E_p^{obs}$ & $\alpha$ & flux & $\log{\frac{L}{erg/s}}$ \\
   &   & (sec) &( keV) & & $(ph/cm^2/s)$&  \\ \hline
\endhead

\hline \multicolumn{7}{c}{{Continued on next page}} \\ \hline
\endfoot

\hline \hline
\endlastfoot
  \hline
GRB050126&  1.29&   24.80&$ 131.10\pm{ 131.10}$ &$-0.74\pm{ 0.72}$ &$ 0.71\pm{  0.17}$ &$49.78_{- 0.39}^{+ 1.10}$\\
GRB050223&  0.59&   22.50&$ 234.70\pm{ 234.70}$ &$-1.78\pm{ 0.74}$ &$ 0.69\pm{  0.16}$ &$49.61_{- 0.52}^{+ 1.42}$\\
GRB050315&  1.95&   95.60&$  32.20\pm{  24.30}$ &$-1.53\pm{ 0.66}$ &$ 1.93\pm{  0.22}$ &$50.20_{- 0.29}^{+ 0.32}$\\
GRB050318&  1.44&   32.00&$  70.30\pm{  20.90}$ &$-0.88\pm{ 0.46}$ &$ 3.16\pm{  0.20}$ &$50.30_{- 0.19}^{+ 0.38}$\\
GRB050319&  3.24&  152.50&$  36.20\pm{  36.20}$ &$-1.46\pm{ 0.96}$ &$ 1.52\pm{  0.21}$ &$50.08_{- 0.30}^{+ 0.34}$\\
GRB050401&  2.90&   33.30&$  99.10\pm{  30.00}$ &$0.04\pm{ 0.70}$ &$10.70\pm{  0.92}$ &$50.96_{- 0.32}^{+ 0.45}$\\
GRB050406&  2.44&    5.40&$  26.20\pm{  18.60}$ &$-0.05\pm{ 2.50}$ &$ 0.36\pm{  0.10}$ &$49.25_{- 0.33}^{+ 0.55}$\\
GRB050416A&  0.65&    2.50&$  13.00\pm{  10.30}$ &$-1.23\pm{ 1.20}$ &$ 4.88\pm{  0.48}$ &$50.31_{- 0.39}^{+ 0.24}$\\
GRB050505&  4.27&   58.90&$ 133.80\pm{ 133.80}$ &$-0.99\pm{ 0.51}$ &$ 1.85\pm{  0.31}$ &$50.24_{- 0.35}^{+ 0.98}$\\
GRB050525A&  0.61&    8.80&$ 109.00\pm{   9.00}$ &$-0.78\pm{ 0.14}$ &$41.70\pm{  0.94}$ &$51.20_{- 0.11}^{+ 0.08}$\\
GRB050802&  1.71&   19.00&$ 140.00\pm{ 140.00}$ &$-0.83\pm{ 1.00}$ &$ 2.75\pm{  0.44}$ &$50.44_{- 0.44}^{+ 1.50}$\\
GRB050814&  5.30&  150.90&$  58.00\pm{  58.00}$ &$-1.19\pm{ 0.74}$ &$ 0.71\pm{  0.25}$ &$49.64_{- 0.23}^{+ 0.54}$\\
GRB050908&  3.35&   19.40&$  49.90\pm{  59.70}$ &$-1.16\pm{ 0.75}$ &$ 0.70\pm{  0.14}$ &$49.68_{- 0.23}^{+ 0.58}$\\
GRB050915A&  2.53&   52.00&$ 122.60\pm{ 122.60}$ &$-0.96\pm{ 0.61}$ &$ 0.77\pm{  0.14}$ &$49.88_{- 0.33}^{+ 1.02}$\\
GRB050922C&  2.20&    4.50&$ 153.10\pm{ 138.90}$ &$-0.91\pm{ 0.30}$ &$ 7.26\pm{  0.32}$ &$50.91_{- 0.36}^{+ 0.70}$\\
GRB051001&  2.43&  189.10&$  68.20\pm{  68.20}$ &$-0.78\pm{ 1.77}$ &$ 0.49\pm{  0.11}$ &$49.53_{- 0.30}^{+ 0.96}$\\
GRB051006&  1.06&   34.80&$ 118.20\pm{ 118.20}$ &$-0.52\pm{ 1.27}$ &$ 1.62\pm{  0.30}$ &$50.05_{- 0.40}^{+ 2.03}$\\
GRB051016B&  0.94&    4.00&$  15.00\pm{  15.30}$ &$-1.74\pm{ 0.66}$ &$ 1.30\pm{  0.16}$ &$49.94_{- 0.34}^{+ 0.21}$\\
GRB051109A&  2.35&   37.20&$  84.00\pm{  84.00}$ &$-0.59\pm{ 1.22}$ &$ 3.94\pm{  0.69}$ &$50.47_{- 0.31}^{+ 1.08}$\\
GRB051109B&  0.08&   14.30&$  53.60\pm{  53.60}$ &$-1.96\pm{ 0.99}$ &$ 0.55\pm{  0.13}$ &$48.09_{- 0.52}^{+ 0.38}$\\
GRB051117B&  0.48&    9.00&$  62.80\pm{  43.90}$ &$1.70\pm{ 2.05}$ &$ 0.49\pm{  0.14}$ &$48.97_{- 1.00}^{+ 0.86}$\\
GRB060108&  2.03&   14.30&$ 196.00\pm{ 196.00}$ &$-2.03\pm{ 0.53}$ &$ 0.77\pm{  0.12}$ &$50.13_{- 0.63}^{+ 1.18}$\\
GRB060124&  2.30&   13.42&$  74.80\pm{  74.80}$ &$-1.09\pm{ 1.38}$ &$ 0.89\pm{  0.18}$ &$49.84_{- 0.32}^{+ 1.28}$\\
GRB060206&  4.04&    7.60&$  74.00\pm{  17.00}$ &$-0.75\pm{ 0.43}$ &$ 2.79\pm{  0.17}$ &$50.27_{- 0.18}^{+ 0.39}$\\
GRB060210&  3.91&  255.00&$ 164.20\pm{ 164.20}$ &$-1.28\pm{ 0.33}$ &$ 2.72\pm{  0.28}$ &$50.47_{- 0.37}^{+ 0.66}$\\
GRB060218&  0.03& 2100.00&$  29.10\pm{  18.50}$ &$-0.20\pm{ 2.17}$ &$ 0.25\pm{  0.11}$ &$46.58_{- 0.92}^{+ 0.96}$\\
GRB060223A&  4.41&   11.30&$  72.00\pm{ 121.80}$ &$-1.22\pm{ 0.52}$ &$ 1.35\pm{  0.18}$ &$49.99_{- 0.23}^{+ 0.85}$\\
GRB060502A&  1.51&   28.40&$ 324.70\pm{ 324.70}$ &$-0.73\pm{ 0.71}$ &$ 1.69\pm{  0.21}$ &$50.55_{- 0.74}^{+ 1.80}$\\
GRB060505&  0.09&    4.00&$  99.30\pm{  99.30}$ &$-0.27\pm{ 3.15}$ &$ 2.65\pm{  0.63}$ &$48.65_{- 1.35}^{+ 2.04}$\\
GRB060512&  0.44&    8.50&$  23.00\pm{  16.00}$ &$-1.01\pm{ 1.75}$ &$ 0.88\pm{  0.20}$ &$49.21_{- 0.65}^{+ 0.77}$\\
GRB060522&  5.11&   71.10&$  75.60\pm{ 110.70}$ &$-0.80\pm{ 0.76}$ &$ 0.55\pm{  0.15}$ &$49.53_{- 0.25}^{+ 1.17}$\\
GRB060526&  3.21&  298.20&$  84.80\pm{  59.60}$ &$-0.25\pm{ 0.92}$ &$ 1.67\pm{  0.18}$ &$50.09_{- 0.29}^{+ 1.06}$\\
GRB060602A&  0.79&   75.00&$ 108.70\pm{ 108.70}$ &$1.31\pm{ 3.65}$ &$ 0.56\pm{  0.20}$ &$49.51_{- 1.36}^{+ 2.79}$\\
GRB060604&  2.14&   95.00&$ 177.40\pm{ 177.40}$ &$-1.65\pm{ 0.74}$ &$ 0.34\pm{  0.13}$ &$49.64_{- 0.47}^{+ 0.96}$\\
GRB060614&  0.13&  108.70&$ 245.10\pm{ 245.10}$ &$-1.52\pm{ 0.32}$ &$11.50\pm{  0.74}$ &$49.80_{- 0.39}^{+ 0.69}$\\
GRB060707&  3.43&   66.20&$  63.40\pm{  15.40}$ &$-0.64\pm{ 0.65}$ &$ 1.01\pm{  0.23}$ &$49.81_{- 0.17}^{+ 0.43}$\\
GRB060714&  2.71&  115.00&$  51.00\pm{  51.00}$ &$-1.77\pm{ 0.74}$ &$ 1.28\pm{  0.13}$ &$50.16_{- 0.40}^{+ 0.23}$\\
GRB060719&  1.53&   66.90&$  51.60\pm{  10.00}$ &$-0.62\pm{ 0.72}$ &$ 2.16\pm{  0.20}$ &$50.07_{- 0.15}^{+ 0.45}$\\
GRB060729&  0.54&  115.30&$ 109.00\pm{ 109.00}$ &$-1.78\pm{ 0.54}$ &$ 1.17\pm{  0.13}$ &$49.72_{- 0.36}^{+ 0.58}$\\
GRB060814&  0.84&  145.30&$ 182.00\pm{ 133.80}$ &$-1.34\pm{ 0.15}$ &$ 7.27\pm{  0.29}$ &$50.72_{- 0.29}^{+ 0.39}$\\
GRB060904B&  0.70&  171.50&$ 315.20\pm{ 315.20}$ &$-1.02\pm{ 0.48}$ &$ 2.44\pm{  0.21}$ &$50.37_{- 0.60}^{+ 1.21}$\\
GRB060906&  3.69&   43.50&$  48.70\pm{  38.70}$ &$-1.53\pm{ 0.46}$ &$ 1.97\pm{  0.28}$ &$50.22_{- 0.25}^{+ 0.31}$\\
GRB060908&  1.88&   19.30&$ 150.70\pm{ 112.40}$ &$-0.97\pm{ 0.27}$ &$ 3.03\pm{  0.25}$ &$50.51_{- 0.34}^{+ 0.63}$\\
GRB060912A&  0.94&    5.00&$  92.80\pm{ 118.70}$ &$-1.30\pm{ 0.34}$ &$ 8.58\pm{  0.44}$ &$50.70_{- 0.24}^{+ 0.68}$\\
GRB060926&  3.21&    8.00&$  15.50\pm{  16.00}$ &$-1.69\pm{ 1.03}$ &$ 1.09\pm{  0.14}$ &$50.06_{- 0.42}^{+ 0.21}$\\
GRB060927&  5.60&   22.50&$  72.00\pm{  17.65}$ &$-0.92\pm{ 0.38}$ &$ 2.70\pm{  0.17}$ &$50.20_{- 0.17}^{+ 0.36}$\\
GRB061110A&  0.76&   40.70&$ 322.50\pm{ 322.50}$ &$-1.61\pm{ 0.41}$ &$ 0.53\pm{  0.12}$ &$49.64_{- 0.45}^{+ 1.00}$\\
GRB061210&  0.41&   85.30&$ 328.90\pm{ 328.90}$ &$-0.51\pm{ 0.57}$ &$ 5.31\pm{  0.47}$ &$50.57_{- 0.81}^{+ 1.66}$\\
GRB061222A&  2.09&   71.40&$ 210.70\pm{  74.10}$ &$-0.88\pm{ 0.15}$ &$ 8.53\pm{  0.26}$ &$51.09_{- 0.41}^{+ 0.39}$\\
GRB061222B&  3.35&   40.00&$  29.10\pm{  21.10}$ &$-0.64\pm{ 2.49}$ &$ 1.59\pm{  0.36}$ &$49.94_{- 0.40}^{+ 0.49}$\\
GRB070103&  2.62&   18.60&$  43.60\pm{  10.70}$ &$0.11\pm{ 1.60}$ &$ 1.04\pm{  0.15}$ &$49.73_{- 0.16}^{+ 0.53}$\\
GRB070129&  2.34&  460.60&$  43.50\pm{  43.50}$ &$-1.45\pm{ 0.74}$ &$ 0.55\pm{  0.12}$ &$49.65_{- 0.29}^{+ 0.39}$\\
GRB070208&  1.17&   47.70&$  47.30\pm{  47.30}$ &$-1.38\pm{ 0.74}$ &$ 0.90\pm{  0.22}$ &$49.73_{- 0.27}^{+ 0.35}$\\
GRB070306&  1.50&  209.50&$ 306.00\pm{ 306.00}$ &$-1.67\pm{ 0.74}$ &$ 4.07\pm{  0.21}$ &$50.75_{- 0.52}^{+ 1.37}$\\
GRB070318&  0.84&   74.60&$ 249.70\pm{ 249.70}$ &$-1.26\pm{ 0.26}$ &$ 1.76\pm{  0.15}$ &$50.19_{- 0.42}^{+ 0.71}$\\
GRB070411&  2.95&  121.50&$ 139.10\pm{ 139.10}$ &$-1.34\pm{ 1.01}$ &$ 0.91\pm{  0.13}$ &$49.99_{- 0.40}^{+ 1.42}$\\
GRB070506&  2.31&    4.30&$  47.30\pm{  12.00}$ &$0.04\pm{ 1.23}$ &$ 0.96\pm{  0.13}$ &$49.71_{- 0.17}^{+ 0.52}$\\
GRB070508&  0.82&   20.90&$ 189.70\pm{  56.40}$ &$-0.64\pm{ 0.18}$ &$24.10\pm{  0.61}$ &$51.30_{- 0.44}^{+ 0.41}$\\
GRB070521&  0.55&   37.90&$ 158.40\pm{  68.90}$ &$-0.55\pm{ 0.29}$ &$ 6.53\pm{  0.27}$ &$50.48_{- 0.41}^{+ 0.53}$\\
GRB070529&  2.50&  109.20&$ 229.60\pm{ 229.60}$ &$-1.18\pm{ 0.45}$ &$ 1.43\pm{  0.36}$ &$50.31_{- 0.48}^{+ 1.13}$\\
GRB070611&  2.04&   12.20&$  57.70\pm{  57.70}$ &$-0.37\pm{ 0.74}$ &$ 0.82\pm{  0.21}$ &$49.69_{- 0.21}^{+ 0.86}$\\
GRB070612A&  0.62&  368.80&$ 171.30\pm{ 171.30}$ &$-0.87\pm{ 1.42}$ &$ 1.51\pm{  0.38}$ &$49.91_{- 0.52}^{+ 2.30}$\\
GRB070714B&  0.92&   64.00&$ 129.60\pm{ 129.60}$ &$-0.96\pm{ 0.68}$ &$ 2.70\pm{  0.20}$ &$50.24_{- 0.38}^{+ 1.08}$\\
GRB070721B&  3.63&  340.00&$ 184.40\pm{ 184.40}$ &$-0.53\pm{ 1.07}$ &$ 1.50\pm{  0.30}$ &$50.32_{- 0.55}^{+ 1.94}$\\
GRB070802&  2.45&   16.40&$  96.50\pm{  96.50}$ &$-2.22\pm{ 0.74}$ &$ 0.40\pm{  0.10}$ &$49.85_{- 0.62}^{+ 0.68}$\\
GRB070810A&  2.17&   11.00&$  52.40\pm{  52.40}$ &$-1.17\pm{ 0.88}$ &$ 1.90\pm{  0.20}$ &$50.13_{- 0.25}^{+ 0.56}$\\
GRB071010A&  0.98&    6.00&$  35.60\pm{  35.60}$ &$-0.69\pm{ 0.74}$ &$ 0.80\pm{  0.30}$ &$49.49_{- 0.15}^{+ 0.47}$\\
GRB071010B&  0.95&   36.12&$  52.20\pm{   6.40}$ &$-1.52\pm{ 0.22}$ &$ 7.70\pm{  0.30}$ &$50.64_{- 0.16}^{+ 0.29}$\\
GRB071028B&  0.94&   55.00&$ 268.00\pm{  49.80}$ &$-2.80\pm{ 0.74}$ &$ 1.40\pm{  0.50}$ &$50.22_{- 0.34}^{+ 0.72}$\\
GRB071117&  1.33&    6.60&$ 127.20\pm{  94.00}$ &$-1.23\pm{ 0.24}$ &$11.30\pm{  0.40}$ &$50.98_{- 0.29}^{+ 0.52}$\\
GRB071122&  1.14&   68.70&$  79.40\pm{  79.40}$ &$-1.42\pm{ 0.94}$ &$ 0.40\pm{  0.20}$ &$49.44_{- 0.32}^{+ 0.70}$\\
GRB080207&  2.09&  340.00&$ 271.00\pm{ 271.00}$ &$-0.84\pm{ 2.18}$ &$ 1.00\pm{  0.30}$ &$50.26_{- 0.73}^{+ 4.43}$\\
GRB080310&  2.43&  365.00&$  41.90\pm{  17.50}$ &$-0.67\pm{ 1.45}$ &$ 1.30\pm{  0.20}$ &$49.88_{- 0.21}^{+ 0.48}$\\
GRB080319C&  1.95&   34.00&$ 244.00\pm{ 244.00}$ &$-0.86\pm{ 0.41}$ &$ 5.20\pm{  0.30}$ &$50.93_{- 0.58}^{+ 1.09}$\\
GRB080413A&  2.43&   46.00&$ 160.30\pm{ 111.60}$ &$-0.73\pm{ 0.30}$ &$ 5.60\pm{  0.20}$ &$50.83_{- 0.39}^{+ 0.81}$\\
GRB080413B&  1.10&    8.00&$ 102.00\pm{  35.50}$ &$-1.00\pm{ 0.29}$ &$18.70\pm{  0.80}$ &$51.09_{- 0.23}^{+ 0.31}$\\
GRB080430&  0.75&   16.20&$ 268.00\pm{ 268.00}$ &$-1.75\pm{ 0.43}$ &$ 2.60\pm{  0.20}$ &$50.31_{- 0.42}^{+ 0.79}$\\
GRB080516&  3.20&    5.80&$  69.10\pm{ 222.80}$ &$-0.35\pm{ 1.35}$ &$ 1.80\pm{  0.30}$ &$50.07_{- 0.24}^{+ 2.58}$\\
GRB080520&  1.54&    2.80&$  28.70\pm{  13.10}$ &$-0.42\pm{ 2.50}$ &$ 0.50\pm{  0.10}$ &$49.38_{- 0.49}^{+ 0.50}$\\
GRB080605&  1.64&   20.00&$ 257.60\pm{ 226.30}$ &$-0.68\pm{ 0.22}$ &$19.90\pm{  0.60}$ &$51.55_{- 0.58}^{+ 0.96}$\\
GRB080607&  3.04&   79.00&$ 326.60\pm{ 326.60}$ &$-0.62\pm{ 0.32}$ &$23.10\pm{  1.10}$ &$51.77_{- 0.71}^{+ 1.23}$\\
GRB080707&  1.23&   27.10&$  27.80\pm{  27.80}$ &$-1.34\pm{ 1.30}$ &$ 1.00\pm{  0.10}$ &$49.77_{- 0.31}^{+ 0.36}$\\
GRB080710&  0.84&  120.00&$  49.00\pm{  49.00}$ &$-1.13\pm{ 1.74}$ &$ 1.00\pm{  0.20}$ &$49.61_{- 0.31}^{+ 0.80}$\\
GRB080804&  2.21&   34.00&$ 143.90\pm{ 143.90}$ &$-1.02\pm{ 0.82}$ &$ 3.10\pm{  0.40}$ &$50.52_{- 0.40}^{+ 1.13}$\\
GRB080805&  1.50&   78.00&$ 135.30\pm{ 319.50}$ &$0.29\pm{ 1.06}$ &$ 1.10\pm{  0.10}$ &$50.08_{- 0.51}^{+ 3.30}$\\
GRB080906&  2.00&  147.00&$  81.20\pm{  81.20}$ &$-0.81\pm{ 1.33}$ &$ 1.00\pm{  0.20}$ &$49.87_{- 0.30}^{+ 1.05}$\\
GRB080913&  6.44&    8.00&$  78.40\pm{  78.40}$ &$-0.58\pm{ 1.33}$ &$ 1.40\pm{  0.20}$ &$49.88_{- 0.29}^{+ 1.04}$\\
GRB080916A&  0.69&   60.00&$ 108.70\pm{  30.30}$ &$-0.03\pm{ 0.47}$ &$ 2.70\pm{  0.20}$ &$50.09_{- 0.33}^{+ 0.46}$\\
GRB080928&  1.69&  280.00&$ 148.60\pm{ 148.60}$ &$-1.25\pm{ 0.38}$ &$ 2.10\pm{  0.10}$ &$50.34_{- 0.33}^{+ 0.72}$\\
GRB081008&  1.97&  185.50&$ 113.40\pm{ 113.40}$ &$-0.88\pm{ 0.76}$ &$ 1.30\pm{  0.10}$ &$50.07_{- 0.33}^{+ 1.20}$\\
GRB081118&  2.58&   67.00&$  72.30\pm{  72.30}$ &$-1.25\pm{ 1.72}$ &$ 0.60\pm{  0.20}$ &$49.69_{- 0.36}^{+ 0.96}$\\
GRB081121&  2.51&   14.00&$  72.30\pm{  72.30}$ &$-1.25\pm{ 1.72}$ &$ 4.40\pm{  1.00}$ &$50.55_{- 0.35}^{+ 0.95}$\\
GRB081203A&  2.10&  294.00&$ 111.30\pm{  85.40}$ &$-0.63\pm{ 0.51}$ &$ 2.90\pm{  0.20}$ &$50.42_{- 0.32}^{+ 0.78}$\\
GRB081221&  2.26&   34.00&$ 112.90\pm{  19.30}$ &$-0.75\pm{ 0.19}$ &$18.20\pm{  0.50}$ &$51.22_{- 0.23}^{+ 0.17}$\\
GRB081222&  2.70&   24.00&$ 157.50\pm{  51.30}$ &$-0.74\pm{ 0.17}$ &$ 7.70\pm{  0.20}$ &$50.96_{- 0.35}^{+ 0.36}$\\
GRB090102&  1.55&   27.00&$ 101.30\pm{ 105.80}$ &$0.18\pm{ 1.13}$ &$ 5.50\pm{  0.80}$ &$50.64_{- 0.36}^{+ 1.50}$\\
GRB090205&  4.70&    8.80&$  33.00\pm{  33.00}$ &$-0.54\pm{ 0.74}$ &$ 0.50\pm{  0.10}$ &$49.38_{- 0.16}^{+ 0.51}$\\
GRB090407&  1.45&  310.00&$ 153.20\pm{ 153.20}$ &$-0.38\pm{ 1.26}$ &$ 0.60\pm{  0.10}$ &$49.81_{- 0.52}^{+ 1.80}$\\
GRB090423&  8.00&   10.30&$  84.50\pm{  28.60}$ &$-1.20\pm{ 0.52}$ &$ 1.70\pm{  0.20}$ &$49.96_{- 0.28}^{+ 0.33}$\\
GRB090424&  0.54&   48.00&$ 166.00\pm{  52.60}$ &$-0.92\pm{ 0.14}$ &$71.00\pm{  2.00}$ &$51.50_{- 0.33}^{+ 0.30}$\\
GRB090529&  2.62&   10.40&$  50.00\pm{  19.30}$ &$0.59\pm{ 1.04}$ &$ 0.40\pm{  0.10}$ &$49.32_{- 0.15}^{+ 0.52}$\\
GRB090530&  1.27&   48.00&$  70.60\pm{  43.30}$ &$-0.36\pm{ 0.95}$ &$ 2.50\pm{  0.30}$ &$50.13_{- 0.23}^{+ 0.84}$\\
GRB090618&  0.54&  113.20&$ 170.40\pm{  68.70}$ &$-1.15\pm{ 0.14}$ &$38.90\pm{  0.80}$ &$51.23_{- 0.30}^{+ 0.32}$\\
GRB090715B&  3.00&  266.00&$ 357.00\pm{ 357.00}$ &$-1.18\pm{ 0.25}$ &$ 3.80\pm{  0.20}$ &$50.86_{- 0.55}^{+ 0.86}$\\
GRB090726&  2.71&   67.00&$  40.30\pm{   9.00}$ &$1.93\pm{ 0.28}$ &$ 0.70\pm{  0.20}$ &$49.49_{- 0.08}^{+ 0.49}$\\
GRB090926B&  1.24&  109.70&$ 177.20\pm{ 177.20}$ &$-0.82\pm{ 0.59}$ &$ 3.20\pm{  0.30}$ &$50.52_{- 0.46}^{+ 1.35}$\\
GRB090927&  1.37&    2.20&$  56.00\pm{  36.40}$ &$-0.68\pm{ 0.98}$ &$ 2.00\pm{  0.20}$ &$50.03_{- 0.20}^{+ 0.65}$\\
GRB091018&  0.97&    4.40&$  35.20\pm{   5.00}$ &$-1.30\pm{ 0.31}$ &$10.30\pm{  0.40}$ &$50.70_{- 0.16}^{+ 0.34}$\\
GRB091029&  2.75&   39.20&$  52.80\pm{  17.70}$ &$-0.92\pm{ 0.65}$ &$ 1.80\pm{  0.10}$ &$50.07_{- 0.19}^{+ 0.42}$\\
GRB091109A&  3.50&   48.00&$ 339.80\pm{ 339.80}$ &$-3.24\pm{ 0.74}$ &$ 1.30\pm{  0.40}$ &$50.36_{- 0.62}^{+ 2.20}$\\
GRB091127&  0.49&    7.10&$  69.50\pm{  34.20}$ &$-1.33\pm{ 0.40}$ &$46.50\pm{  2.70}$ &$51.09_{- 0.20}^{+ 0.33}$\\
GRB091208B&  1.06&   14.90&$ 255.50\pm{ 255.50}$ &$-1.31\pm{ 0.35}$ &$15.20\pm{  1.00}$ &$51.21_{- 0.45}^{+ 1.07}$\\
GRB100302A&  4.81&   17.90&$  68.00\pm{  68.00}$ &$-1.49\pm{ 0.74}$ &$ 0.50\pm{  0.10}$ &$49.60_{- 0.32}^{+ 0.50}$\\
GRB100316B&  1.18&    3.80&$  30.20\pm{  21.70}$ &$-1.53\pm{ 0.73}$ &$ 1.30\pm{  0.10}$ &$49.93_{- 0.30}^{+ 0.32}$\\
GRB100418A&  0.62&    7.00&$  18.50\pm{  18.50}$ &$-1.98\pm{ 0.74}$ &$ 1.00\pm{  0.20}$ &$49.79_{- 0.50}^{+ 0.07}$\\
GRB100425A&  1.75&   37.00&$  26.60\pm{  26.60}$ &$-0.85\pm{ 1.67}$ &$ 1.40\pm{  0.20}$ &$49.90_{- 0.25}^{+ 0.45}$\\
GRB100615A&  1.40&   39.00&$  94.80\pm{  43.50}$ &$-1.16\pm{ 0.24}$ &$ 5.40\pm{  0.20}$ &$50.61_{- 0.20}^{+ 0.29}$\\
GRB100621A&  0.54&   63.60&$  89.00\pm{  11.60}$ &$-0.92\pm{ 0.14}$ &$12.80\pm{  0.30}$ &$50.58_{- 0.12}^{+ 0.11}$\\
GRB100704A&  3.60&  197.50&$ 179.30\pm{ 250.70}$ &$-0.59\pm{ 0.33}$ &$ 4.30\pm{  0.20}$ &$50.76_{- 0.46}^{+ 1.29}$\\
GRB100728B&  2.80&   12.10&$  71.20\pm{  38.20}$ &$-0.10\pm{ 1.25}$ &$ 3.50\pm{  0.50}$ &$50.37_{- 0.26}^{+ 0.60}$\\
GRB100814A&  1.44&  174.50&$ 183.00\pm{  82.00}$ &$0.30\pm{ 0.70}$ &$ 2.50\pm{  0.20}$ &$50.60_{- 0.66}^{+ 1.11}$\\
GRB100816A&  0.80&    2.90&$ 147.80\pm{  44.00}$ &$-0.45\pm{ 0.24}$ &$10.90\pm{  0.40}$ &$50.87_{- 0.38}^{+ 0.44}$\\
GRB100902A&  4.50&  428.80&$  63.60\pm{  63.60}$ &$-1.09\pm{ 1.14}$ &$ 1.00\pm{  0.10}$ &$49.81_{- 0.27}^{+ 0.85}$\\
GRB100906A&  1.73&  114.40&$  97.90\pm{  26.90}$ &$-0.88\pm{ 0.29}$ &$10.10\pm{  0.40}$ &$50.91_{- 0.23}^{+ 0.25}$\\
GRB101219B&  0.55&   34.00&$  65.70\pm{  65.70}$ &$2.10\pm{ 0.59}$ &$ 0.60\pm{  0.30}$ &$49.16_{- 0.25}^{+ 1.22}$\\
GRB110128A&  2.34&   30.70&$  92.50\pm{  92.50}$ &$-0.05\pm{ 2.24}$ &$ 0.80\pm{  0.20}$ &$49.81_{- 0.58}^{+ 1.70}$\\
GRB110205A&  1.98&  257.00&$ 144.60\pm{  46.00}$ &$-1.22\pm{ 0.34}$ &$ 3.60\pm{  0.20}$ &$50.58_{- 0.31}^{+ 0.34}$\\
GRB110213A&  1.46&   48.00&$  23.70\pm{  23.70}$ &$-1.34\pm{ 0.74}$ &$ 1.60\pm{  0.60}$ &$50.03_{- 0.24}^{+ 0.35}$\\
GRB110422A&  1.77&   25.90&$ 281.60\pm{ 281.60}$ &$-0.70\pm{ 0.22}$ &$30.70\pm{  1.00}$ &$51.78_{- 0.60}^{+ 0.96}$\\
GRB110503A&  1.61&   10.00&$ 111.80\pm{  21.90}$ &$-0.10\pm{ 0.36}$ &$ 1.35\pm{  0.06}$ &$50.07_{- 0.28}^{+ 0.32}$\\
GRB110715A&  0.82&   13.00&$ 152.00\pm{  32.60}$ &$-0.98\pm{ 0.13}$ &$53.90\pm{  1.10}$ &$51.55_{- 0.26}^{+ 0.20}$\\
GRB110731A&  2.83&   38.80&$ 135.20\pm{  39.30}$ &$-0.77\pm{ 0.22}$ &$11.00\pm{  0.30}$ &$51.06_{- 0.31}^{+ 0.38}$\\
GRB110801A&  1.86&  385.00&$  62.50\pm{  62.50}$ &$-1.65\pm{ 0.94}$ &$ 1.10\pm{  0.20}$ &$50.02_{- 0.38}^{+ 0.53}$\\
GRB110808A&  1.35&   48.00&$  65.80\pm{  65.80}$ &$-0.17\pm{ 1.11}$ &$ 0.40\pm{  0.20}$ &$49.33_{- 0.23}^{+ 0.88}$\\
GRB111008A&  5.00&   63.46&$ 212.30\pm{ 212.30}$ &$-1.01\pm{ 0.54}$ &$ 6.40\pm{  0.70}$ &$50.89_{- 0.48}^{+ 1.34}$\\
GRB111209A&  0.68&  810.96&$ 118.60\pm{ 118.60}$ &$-1.57\pm{ 0.77}$ &$ 0.50\pm{  0.10}$ &$49.41_{- 0.42}^{+ 0.86}$\\
GRB111225A&  0.30&  106.80&$  40.20\pm{  48.10}$ &$0.70\pm{ 4.08}$ &$ 0.70\pm{  0.10}$ &$48.72_{- 1.81}^{+ 1.13}$\\
GRB111228A&  0.71&  101.20&$  88.70\pm{  29.70}$ &$-1.65\pm{ 0.27}$ &$12.40\pm{  0.50}$ &$50.82_{- 0.23}^{+ 0.22}$\\
GRB111229A&  1.38&   25.40&$ 102.70\pm{ 102.70}$ &$-1.38\pm{ 0.90}$ &$ 1.00\pm{  0.20}$ &$49.91_{- 0.36}^{+ 0.82}$\\
GRB120118B&  2.94&   23.26&$  50.70\pm{  50.70}$ &$-1.43\pm{ 1.08}$ &$ 2.20\pm{  0.30}$ &$50.26_{- 0.31}^{+ 0.54}$\\
GRB120119A&  1.73&  253.80&$ 249.60\pm{ 249.60}$ &$-0.86\pm{ 0.21}$ &$10.30\pm{  0.30}$ &$51.22_{- 0.52}^{+ 0.92}$\\
GRB120326A&  1.80&   69.60&$  48.40\pm{   6.90}$ &$-1.13\pm{ 0.41}$ &$ 4.60\pm{  0.20}$ &$50.48_{- 0.15}^{+ 0.39}$\\
GRB120404A&  2.88&   38.70&$  21.50\pm{  21.50}$ &$-2.05\pm{ 0.74}$ &$ 1.20\pm{  0.20}$ &$50.30_{- 0.60}^{+ 0.02}$\\
GRB120422A&  0.28&    5.35&$ 103.60\pm{  34.00}$ &$0.56\pm{ 2.95}$ &$ 0.60\pm{  0.20}$ &$48.91_{- 1.14}^{+ 0.90}$\\
GRB120521C&  6.00&   26.70&$ 361.80\pm{ 361.80}$ &$-1.15\pm{ 0.50}$ &$ 1.90\pm{  0.20}$ &$50.45_{- 0.61}^{+ 1.60}$\\
GRB120712A&  4.00&   14.70&$  99.70\pm{  55.80}$ &$-0.13\pm{ 0.74}$ &$ 2.40\pm{  0.20}$ &$50.28_{- 0.33}^{+ 0.68}$\\
GRB120722A&  0.96&   42.40&$  69.70\pm{  69.70}$ &$-2.17\pm{ 0.74}$ &$ 1.00\pm{  0.30}$ &$50.06_{- 0.63}^{+ 0.32}$\\
GRB120724A&  1.48&   72.80&$  41.10\pm{  10.00}$ &$2.91\pm{ 0.74}$ &$ 0.60\pm{  0.20}$ &$49.37_{- 0.26}^{+ 0.50}$\\
GRB120729A&  0.80&   71.50&$  64.80\pm{  17.50}$ &$-0.78\pm{ 0.52}$ &$ 2.90\pm{  0.20}$ &$50.05_{- 0.18}^{+ 0.40}$\\
GRB120802A&  3.80&   50.00&$  84.10\pm{  84.10}$ &$-1.19\pm{ 0.50}$ &$ 3.00\pm{  0.20}$ &$50.38_{- 0.24}^{+ 0.64}$\\
GRB120805A&  3.10&   48.00&$ 307.10\pm{ 307.10}$ &$-3.07\pm{ 0.74}$ &$ 0.37\pm{  0.20}$ &$49.82_{- 0.63}^{+ 2.22}$\\
GRB120811C&  2.67&   26.80&$  53.90\pm{   9.80}$ &$-1.01\pm{ 0.46}$ &$ 4.10\pm{  0.20}$ &$50.45_{- 0.15}^{+ 0.41}$\\
GRB120815A&  2.36&    9.70&$  45.70\pm{  45.70}$ &$-1.22\pm{ 1.09}$ &$ 2.20\pm{  0.30}$ &$50.20_{- 0.26}^{+ 0.51}$\\
GRB120909A&  3.93&  220.60&$  97.90\pm{ 100.70}$ &$0.41\pm{ 1.43}$ &$ 1.80\pm{  0.30}$ &$50.16_{- 0.41}^{+ 2.03}$\\
GRB121128A&  2.20&   23.30&$ 107.70\pm{  16.20}$ &$-0.49\pm{ 0.23}$ &$12.90\pm{  0.40}$ &$51.06_{- 0.19}^{+ 0.18}$\\
GRB121201A&  3.38&   85.00&$ 115.70\pm{ 115.70}$ &$-1.24\pm{ 1.37}$ &$ 0.80\pm{  0.10}$ &$49.88_{- 0.41}^{+ 1.52}$\\
GRB121211A&  1.02&  182.00&$ 241.60\pm{ 241.60}$ &$-2.87\pm{ 0.74}$ &$ 1.00\pm{  0.30}$ &$50.10_{- 0.63}^{+ 1.47}$\\
GRB130215A&  0.60&   65.70&$  69.90\pm{  69.90}$ &$-0.16\pm{ 2.98}$ &$ 2.50\pm{  0.70}$ &$49.83_{- 1.07}^{+ 1.50}$\\
GRB130408A&  3.76&   28.00&$ 198.40\pm{ 198.40}$ &$-0.98\pm{ 0.30}$ &$ 4.90\pm{  1.00}$ &$50.80_{- 0.44}^{+ 1.01}$\\
GRB130418A&  1.22&  274.92&$  30.70\pm{  12.40}$ &$1.64\pm{ 5.27}$ &$ 0.60\pm{  0.20}$ &$49.29_{- 1.71}^{+ 0.61}$\\
GRB130420A&  1.30&  123.50&$  58.00\pm{  15.60}$ &$-1.03\pm{ 0.49}$ &$ 3.40\pm{  0.20}$ &$50.29_{- 0.18}^{+ 0.40}$\\
GRB130427B&  2.78&   27.00&$ 176.30\pm{ 176.30}$ &$-1.18\pm{ 0.65}$ &$ 3.00\pm{  0.40}$ &$50.56_{- 0.42}^{+ 1.38}$\\
GRB130511A&  1.30&    5.43&$ 210.40\pm{ 210.40}$ &$-1.28\pm{ 1.11}$ &$ 1.30\pm{  0.20}$ &$50.16_{- 0.52}^{+ 1.61}$\\
GRB130514A&  3.60&  204.00&$ 122.90\pm{  40.00}$ &$-1.65\pm{ 0.19}$ &$ 2.80\pm{  0.30}$ &$50.50_{- 0.26}^{+ 0.18}$\\
GRB130604A&  1.06&   37.70&$ 219.10\pm{ 219.10}$ &$-2.72\pm{ 0.74}$ &$ 0.80\pm{  0.20}$ &$50.02_{- 0.61}^{+ 1.50}$\\
GRB130610A&  2.09&   46.40&$  93.70\pm{  65.30}$ &$-0.20\pm{ 0.80}$ &$ 1.70\pm{  0.20}$ &$50.13_{- 0.30}^{+ 0.88}$\\
GRB130612A&  2.01&    4.00&$  29.80\pm{  29.80}$ &$0.96\pm{ 0.74}$ &$ 1.70\pm{  0.30}$ &$49.85_{- 0.06}^{+ 0.59}$\\
GRB130701A&  1.16&    4.38&$  52.60\pm{  52.60}$ &$9.48\pm{ 0.74}$ &$17.10\pm{  0.70}$ &$50.72_{- 1.35}^{+ 1.77}$\\
GRB130831A&  0.48&   32.50&$ 147.70\pm{ 147.70}$ &$-1.54\pm{ 0.24}$ &$13.60\pm{  0.60}$ &$50.69_{- 0.29}^{+ 0.51}$\\
GRB130925A&  0.35&  160.29&$  33.40\pm{  20.00}$ &$-1.85\pm{ 0.14}$ &$ 7.30\pm{  0.60}$ &$50.19_{- 0.16}^{+ 0.20}$\\
GRB131030A&  1.29&   41.10&$ 229.50\pm{ 105.60}$ &$-0.59\pm{ 0.19}$ &$28.10\pm{  0.70}$ &$51.61_{- 0.54}^{+ 0.57}$\\
GRB131103A&  0.60&   17.30&$  76.30\pm{  25.80}$ &$-0.39\pm{ 1.33}$ &$ 1.50\pm{  0.30}$ &$49.64_{- 0.28}^{+ 0.43}$\\
GRB131105A&  1.69&  112.30&$ 116.40\pm{ 116.40}$ &$-0.41\pm{ 1.18}$ &$ 3.50\pm{  0.60}$ &$50.50_{- 0.39}^{+ 1.37}$\\
GRB131117A&  4.18&   11.00&$  53.70\pm{  25.60}$ &$0.17\pm{ 1.87}$ &$ 0.70\pm{  0.10}$ &$49.55_{- 0.23}^{+ 0.50}$\\
GRB131227A&  5.30&   18.00&$ 131.90\pm{ 131.90}$ &$-1.25\pm{ 0.75}$ &$ 1.10\pm{  0.20}$ &$49.98_{- 0.40}^{+ 1.07}$\\
GRB140114A&  3.00&  139.70&$  33.70\pm{  33.70}$ &$-1.63\pm{ 1.23}$ &$ 0.90\pm{  0.10}$ &$49.92_{- 0.39}^{+ 0.28}$\\
GRB140206A&  2.73&   93.60&$  94.50\pm{   8.20}$ &$-0.37\pm{ 0.19}$ &$19.40\pm{  0.50}$ &$51.20_{- 0.11}^{+ 0.10}$\\
GRB140213A&  1.21&   60.00&$ 109.20\pm{ 104.50}$ &$-1.45\pm{ 0.21}$ &$23.50\pm{  0.80}$ &$51.27_{- 0.22}^{+ 0.34}$\\
GRB140301A&  1.42&   31.00&$ 102.50\pm{ 102.50}$ &$-1.56\pm{ 1.53}$ &$ 0.70\pm{  0.20}$ &$49.80_{- 0.45}^{+ 1.21}$\\
GRB140304A&  5.28&   15.60&$ 127.10\pm{  40.30}$ &$-0.72\pm{ 0.65}$ &$ 1.70\pm{  0.20}$ &$50.15_{- 0.35}^{+ 0.55}$\\
GRB140318A&  1.02&    8.43&$ 192.20\pm{ 192.20}$ &$-0.64\pm{ 1.03}$ &$ 0.50\pm{  0.20}$ &$49.71_{- 0.57}^{+ 1.75}$\\
GRB140419A&  3.96&   94.70&$ 172.30\pm{ 195.20}$ &$-0.76\pm{ 0.32}$ &$ 4.90\pm{  0.20}$ &$50.77_{- 0.42}^{+ 1.19}$\\
GRB140423A&  3.26&  134.00&$ 277.80\pm{ 277.80}$ &$-1.00\pm{ 0.61}$ &$ 2.10\pm{  0.20}$ &$50.56_{- 0.59}^{+ 1.43}$\\
GRB140428A&  4.70&   17.42&$  60.00\pm{  22.60}$ &$0.16\pm{ 3.20}$ &$ 0.60\pm{  0.20}$ &$49.49_{- 0.43}^{+ 0.77}$\\
GRB140430A&  1.60&  173.60&$  92.40\pm{ 129.20}$ &$-0.79\pm{ 0.56}$ &$ 2.50\pm{  0.20}$ &$50.27_{- 0.28}^{+ 1.26}$\\
GRB140506A&  0.89&  111.10&$ 100.40\pm{  55.00}$ &$-0.53\pm{ 0.60}$ &$10.90\pm{  0.90}$ &$50.76_{- 0.30}^{+ 0.79}$\\
GRB140512A&  0.72&  154.80&$ 136.70\pm{ 115.50}$ &$-0.99\pm{ 0.30}$ &$ 6.80\pm{  0.30}$ &$50.56_{- 0.32}^{+ 0.68}$\\
GRB140515A&  6.32&   23.40&$  47.10\pm{  26.90}$ &$-0.35\pm{ 1.41}$ &$ 0.90\pm{  0.10}$ &$49.57_{- 0.19}^{+ 0.53}$\\
GRB140518A&  4.71&   60.50&$  48.30\pm{  14.20}$ &$0.10\pm{ 1.51}$ &$ 1.00\pm{  0.10}$ &$49.67_{- 0.18}^{+ 0.52}$\\
GRB140629A&  2.27&   42.00&$ 105.90\pm{ 100.70}$ &$-0.66\pm{ 0.60}$ &$ 4.20\pm{  0.40}$ &$50.57_{- 0.32}^{+ 0.92}$\\
GRB140703A&  3.14&   67.10&$ 101.50\pm{  17.10}$ &$-0.32\pm{ 1.51}$ &$ 2.80\pm{  0.60}$ &$50.38_{- 0.26}^{+ 0.32}$\\
GRB140710A&  0.56&    3.52&$  69.30\pm{  11.10}$ &$-1.20\pm{ 1.02}$ &$ 1.90\pm{  0.30}$ &$49.75_{- 0.23}^{+ 0.35}$\\
GRB140907A&  1.21&   79.20&$ 110.20\pm{ 110.20}$ &$-0.88\pm{ 0.56}$ &$ 2.50\pm{  0.20}$ &$50.26_{- 0.31}^{+ 1.04}$\\
GRB141004A&  0.57&    3.92&$  75.00\pm{  19.40}$ &$-1.11\pm{ 0.30}$ &$ 6.10\pm{  0.30}$ &$50.27_{- 0.18}^{+ 0.34}$\\
GRB141026A&  3.35&  146.00&$ 329.80\pm{ 329.80}$ &$-3.22\pm{ 0.74}$ &$ 0.40\pm{  0.20}$ &$49.85_{- 0.61}^{+ 1.89}$\\
GRB141121A&  1.47&  549.90&$  52.40\pm{  52.40}$ &$-1.34\pm{ 1.95}$ &$ 0.90\pm{  0.30}$ &$49.79_{- 0.43}^{+ 1.06}$\\
GRB141220A&  1.32&    7.21&$ 129.60\pm{  82.10}$ &$-0.27\pm{ 0.56}$ &$ 8.90\pm{  0.70}$ &$50.90_{- 0.42}^{+ 0.82}$\\
GRB150206A&  2.09&   83.20&$ 197.30\pm{ 197.30}$ &$-0.88\pm{ 0.28}$ &$10.10\pm{  0.40}$ &$51.14_{- 0.44}^{+ 0.99}$\\
GRB150314A&  1.76&   14.79&$ 248.10\pm{ 101.30}$ &$-0.47\pm{ 0.17}$ &$38.50\pm{  0.90}$ &$51.87_{- 0.61}^{+ 0.54}$\\
GRB150323A&  0.59&  149.60&$  52.30\pm{  52.30}$ &$-1.77\pm{ 0.36}$ &$ 5.40\pm{  0.30}$ &$50.36_{- 0.27}^{+ 0.23}$\\
GRB150413A&  3.20&  263.60&$  92.00\pm{  54.00}$ &$0.24\pm{ 0.03}$ &$ 1.60\pm{  0.30}$ &$50.10_{- 0.24}^{+ 0.63}$\\
GRB150818A&  0.28&  123.30&$  61.90\pm{  61.90}$ &$-1.91\pm{ 0.68}$ &$ 2.40\pm{  0.30}$ &$49.65_{- 0.44}^{+ 0.19}$\\
GRB150915A&  1.97&  164.70&$  56.50\pm{   2.70}$ &$-0.88\pm{ 3.56}$ &$ 0.50\pm{  0.20}$ &$49.51_{- 0.13}^{+ 0.42}$\\
GRB151029A&  1.42&    8.95&$  40.20\pm{  14.10}$ &$-0.57\pm{ 1.39}$ &$ 1.80\pm{  0.30}$ &$49.94_{- 0.18}^{+ 0.48}$\\
GRB151031A&  1.17&    5.00&$  38.30\pm{  38.30}$ &$-1.40\pm{ 0.89}$ &$ 1.70\pm{  0.20}$ &$50.00_{- 0.26}^{+ 0.36}$\\
GRB151111A&  3.50&   76.93&$  78.40\pm{  38.80}$ &$0.71\pm{ 1.51}$ &$ 1.00\pm{  0.10}$ &$49.84_{- 0.32}^{+ 0.73}$\\
GRB151112A&  4.10&   19.32&$  72.90\pm{  27.30}$ &$-0.95\pm{ 0.94}$ &$ 1.90\pm{  0.20}$ &$50.11_{- 0.25}^{+ 0.43}$\\
GRB151215A&  2.59&   17.80&$  65.30\pm{  18.00}$ &$0.44\pm{ 1.23}$ &$ 1.60\pm{  0.20}$ &$50.00_{- 0.23}^{+ 0.45}$\\
GRB160121A&  1.96&   12.00&$  51.50\pm{  18.10}$ &$-0.93\pm{ 1.19}$ &$ 1.20\pm{  0.20}$ &$49.88_{- 0.25}^{+ 0.43}$\\
GRB160203A&  3.52&   20.20&$  48.70\pm{  48.70}$ &$-2.01\pm{ 2.33}$ &$ 1.30\pm{  0.40}$ &$50.34_{- 0.61}^{+ 0.07}$\\
GRB160227A&  2.38&  316.50&$  76.20\pm{  55.60}$ &$1.22\pm{ 2.44}$ &$ 0.60\pm{  0.10}$ &$49.63_{- 0.37}^{+ 1.10}$\\
GRB160303A&  2.30&    5.00&$ 183.70\pm{ 183.70}$ &$-0.34\pm{ 0.93}$ &$ 1.00\pm{  0.10}$ &$50.18_{- 0.59}^{+ 1.75}$\\
GRB160327A&  4.99&   28.00&$  72.20\pm{  71.20}$ &$-0.85\pm{ 0.69}$ &$ 1.80\pm{  0.20}$ &$50.04_{- 0.22}^{+ 0.72}$\\
GRB160425A&  0.56&  304.58&$ 127.50\pm{ 127.50}$ &$-1.25\pm{ 0.44}$ &$ 2.80\pm{  0.20}$ &$50.03_{- 0.34}^{+ 0.79}$\\
GRB160703A&  1.50&   44.40&$ 164.60\pm{  50.50}$ &$-1.28\pm{ 0.27}$ &$ 5.80\pm{  0.30}$ &$50.78_{- 0.30}^{+ 0.25}$\\
GRB160804A&  0.74&  144.20&$  75.90\pm{  26.30}$ &$-1.16\pm{ 0.77}$ &$ 2.90\pm{  0.30}$ &$50.08_{- 0.27}^{+ 0.35}$\\
GRB161014A&  2.82&   18.30&$  70.30\pm{  70.30}$ &$0.08\pm{ 2.06}$ &$ 2.90\pm{  0.60}$ &$50.28_{- 0.29}^{+ 1.72}$\\
GRB161108A&  1.16&  105.10&$  31.70\pm{  21.20}$ &$-0.43\pm{ 2.58}$ &$ 0.60\pm{  0.10}$ &$49.39_{- 0.84}^{+ 0.51}$\\
GRB161117A&  1.55&  125.70&$  74.40\pm{  16.50}$ &$-1.12\pm{ 0.29}$ &$ 6.80\pm{  0.30}$ &$50.68_{- 0.17}^{+ 0.34}$\\
GRB161129A&  0.65&   35.53&$ 185.70\pm{  54.20}$ &$-0.85\pm{ 0.44}$ &$ 3.40\pm{  0.20}$ &$50.31_{- 0.42}^{+ 0.47}$\\
GRB161219B&  0.15&    6.94&$  67.90\pm{  25.90}$ &$-0.87\pm{ 0.56}$ &$ 5.30\pm{  0.40}$ &$49.27_{- 0.21}^{+ 0.38}$\\
GRB170113A&  1.97&   20.66&$ 139.80\pm{ 139.80}$ &$-0.87\pm{ 0.73}$ &$ 1.10\pm{  0.10}$ &$50.06_{- 0.41}^{+ 1.45}$\\
GRB170202A&  3.65&   46.20&$  73.10\pm{  33.40}$ &$-0.75\pm{ 0.59}$ &$ 4.70\pm{  0.30}$ &$50.51_{- 0.22}^{+ 0.39}$\\
GRB170531B&  2.37&  164.13&$  90.30\pm{  90.30}$ &$-1.61\pm{ 1.55}$ &$ 0.80\pm{  0.20}$ &$49.93_{- 0.48}^{+ 0.93}$\\
GRB170604A&  1.33&   26.70&$  72.90\pm{  22.30}$ &$-0.17\pm{ 1.23}$ &$ 4.20\pm{  0.90}$ &$50.38_{- 0.27}^{+ 0.40}$\\
GRB170705A&  2.01&  217.30&$ 164.10\pm{  93.10}$ &$-0.93\pm{ 0.22}$ &$13.90\pm{  0.40}$ &$51.21_{- 0.36}^{+ 0.58}$\\
GRB170714A&  0.79&  459.22&$  57.80\pm{  20.20}$ &$1.73\pm{ 0.88}$ &$ 0.40\pm{  0.20}$ &$49.10_{- 0.20}^{+ 0.40}$\\
GRB170903A&  0.89&   29.20&$ 181.80\pm{ 181.80}$ &$-1.82\pm{ 0.76}$ &$ 3.90\pm{  0.50}$ &$50.54_{- 0.53}^{+ 0.92}$\\
GRB171020A&  1.87&   41.90&$ 100.00\pm{ 100.00}$ &$-0.79\pm{ 2.63}$ &$ 0.70\pm{  0.20}$ &$49.76_{- 0.56}^{+ 2.09}$\\
GRB171222A&  2.41&  174.80&$  22.40\pm{  22.40}$ &$-1.86\pm{ 1.66}$ &$ 0.70\pm{  0.20}$ &$49.92_{- 0.51}^{+ 0.17}$\\
GRB180115A&  2.49&   40.90&$ 116.90\pm{ 116.90}$ &$-2.33\pm{ 0.65}$ &$ 0.60\pm{  0.20}$ &$50.03_{- 0.62}^{+ 0.92}$\\
GRB180205A&  1.41&   15.50&$  68.40\pm{  57.00}$ &$-1.05\pm{ 0.61}$ &$ 3.40\pm{  0.30}$ &$50.34_{- 0.22}^{+ 0.51}$\\
GRB180314A&  1.45&   51.20&$  62.60\pm{   5.30}$ &$0.83\pm{ 0.66}$ &$ 7.90\pm{  0.60}$ &$50.62_{- 0.10}^{+ 0.14}$\\
GRB180329B&  2.00&  210.00&$ 178.70\pm{ 178.70}$ &$-1.88\pm{ 1.26}$ &$ 1.40\pm{  0.40}$ &$50.33_{- 0.59}^{+ 1.61}$\\
GRB180510B&  1.30&  134.30&$  96.40\pm{  96.40}$ &$-1.38\pm{ 1.18}$ &$ 1.20\pm{  0.20}$ &$49.97_{- 0.40}^{+ 0.98}$\\
GRB180620B&  1.12&  198.80&$ 250.70\pm{ 250.70}$ &$-0.82\pm{ 0.32}$ &$ 3.60\pm{  0.20}$ &$50.67_{- 0.54}^{+ 1.07}$\\
GRB180624A&  2.85&  486.40&$ 128.40\pm{ 128.40}$ &$-1.78\pm{ 0.98}$ &$ 1.40\pm{  0.20}$ &$50.27_{- 0.49}^{+ 1.11}$\\
GRB180720B&  0.65&  108.40&$ 214.70\pm{ 194.20}$ &$-0.66\pm{ 0.26}$ &$67.90\pm{  2.60}$ &$51.70_{- 0.52}^{+ 0.91}$\\
GRB181010A&  1.39&   16.40&$ 297.00\pm{ 297.00}$ &$-1.42\pm{ 0.39}$ &$ 1.40\pm{  0.20}$ &$50.27_{- 0.46}^{+ 1.15}$\\
GRB181110A&  1.50&  138.40&$  62.20\pm{  22.60}$ &$-1.48\pm{ 0.50}$ &$ 3.70\pm{  0.30}$ &$50.46_{- 0.26}^{+ 0.30}$\\
GRB181213A&  2.40&   15.31&$ 341.50\pm{ 341.50}$ &$-1.50\pm{ 1.06}$ &$ 2.10\pm{  0.60}$ &$50.53_{- 0.59}^{+ 2.35}$\\
GRB190106A&  1.86&   76.80&$ 120.10\pm{  82.60}$ &$-1.34\pm{ 0.21}$ &$ 5.50\pm{  0.30}$ &$50.73_{- 0.26}^{+ 0.33}$\\
GRB190114A&  3.38&   66.60&$ 117.90\pm{ 117.90}$ &$-2.33\pm{ 0.67}$ &$ 0.50\pm{  0.20}$ &$49.94_{- 0.62}^{+ 0.88}$\\
GRB190324A&  1.17&   28.40&$  99.60\pm{  30.60}$ &$-0.76\pm{ 0.33}$ &$11.90\pm{  0.50}$ &$50.89_{- 0.25}^{+ 0.28}$\\
GRB190719C&  2.47&  185.70&$ 151.70\pm{  45.60}$ &$-1.08\pm{ 0.37}$ &$ 5.50\pm{  0.30}$ &$50.79_{- 0.36}^{+ 0.37}$\\
GRB190829A&  0.08&   58.20&$  80.00\pm{  80.00}$ &$-1.00\pm{ 0.74}$ &$18.00\pm{  2.70}$ &$49.35_{- 0.29}^{+ 0.95}$\\
GRB191004B&  3.50&   37.70&$ 184.40\pm{ 282.80}$ &$-0.76\pm{ 0.31}$ &$ 5.00\pm{  0.20}$ &$50.82_{- 0.43}^{+ 1.25}$\\
GRB191011A&  1.72&    7.37&$  74.00\pm{  26.30}$ &$-1.16\pm{ 0.67}$ &$ 1.80\pm{  0.20}$ &$50.13_{- 0.24}^{+ 0.35}$\\
GRB191019A&  0.25&   64.35&$  53.50\pm{   9.60}$ &$-0.81\pm{ 0.60}$ &$ 5.70\pm{  0.40}$ &$49.63_{- 0.15}^{+ 0.42}$\\
GRB191221B&  1.19&   48.00&$  57.40\pm{  50.30}$ &$-0.53\pm{ 1.27}$ &$ 4.80\pm{  0.70}$ &$50.37_{- 0.23}^{+ 0.73}$\\
GRB 200205B&  1.465&  458&$ 90.6\pm{ 30}$ &$-1.155\pm{ 0.546}$ &$ 2.1\pm{  0.27}$ &$51.53_{- 0.27}^{+ 0.31}$\\
  \hline
\end{longtable}
}

 \scriptsize {
\begin{longtable}{|c|c|c|c|c|c|c|}
\caption{Data for 49 \textit{Fermi/GBM} GRBs.} \label{tabFermi}\\
  \hline
  GRB & z & $T_{90}$ &  $E_p^{obs}$ & $\alpha$ & flux & $\log{\frac{L}{erg/s}}$ \\
   &   & (sec) & (keV) & &$(ph/cm^2/s)$ & \\
\hline
\endfirsthead
 \multicolumn{7}{c}%
{{\bfseries \tablename\ \thetable{} -- continued from previous page}} \\
\hline  GRB & z & $T_{90}$ &  $E_p^{obs}$ & $\alpha$ & flux & $\log{\frac{L}{erg/s}}$ \\
   &   & (sec) &(keV)  & & $(ph/cm^2/s)$&  \\ \hline
\endhead

\hline \multicolumn{7}{c}{{Continued on next page}} \\ \hline
\endfoot

\hline \hline
\endlastfoot
  \hline
GRB090902B&  1.82&   19.33&$ 1809.62\pm{  87.60}$ &$-1.19\pm{ 0.01}$ &$80.25\pm{  0.60}$ &$53.93_{- 0.04}^{+ 0.04}$\\
GRB091003A&  0.90&   20.22&$ 481.95\pm{  21.23}$ &$-0.69\pm{ 0.03}$ &$34.38\pm{  0.47}$ &$52.62_{- 0.08}^{+ 0.08}$\\
GRB100414A&  1.37&   26.50&$ 522.48\pm{  32.42}$ &$-0.75\pm{ 0.04}$ &$21.33\pm{  0.36}$ &$52.88_{- 0.10}^{+ 0.09}$\\
GRB110402A&  0.81&   35.65&$ 1131.86\pm{ 1131.86}$ &$-1.01\pm{ 0.13}$ &$ 6.66\pm{  0.58}$ &$51.98_{- 0.94}^{+ 0.74}$\\
GRB120712A*&  4.17&   22.53&$ 144.98\pm{  24.88}$ &$-0.33\pm{ 0.31}$ &$ 2.94\pm{  0.26}$ &$52.82_{- 0.30}^{+ 0.23}$\\
GRB120729A&  0.80&   71.46&$ 310.55\pm{  32.00}$ &$-1.61\pm{ 0.08}$ &$ 5.45\pm{  0.44}$ &$51.35_{- 0.06}^{+ 0.06}$\\
GRB120922A*&  3.10&  182.28&$  73.09\pm{  12.26}$ &$-1.18\pm{ 0.35}$ &$ 2.59\pm{  0.26}$ &$52.25_{- 0.18}^{+ 0.27}$\\
GRB130215A&  0.60&  143.75&$ 396.06\pm{ 161.30}$ &$-1.42\pm{ 0.08}$ &$ 3.31\pm{  0.27}$ &$50.90_{- 0.40}^{+ 0.23}$\\
GRB130420A*&  1.30&  104.96&$  58.71\pm{   5.75}$ &$-0.03\pm{ 0.49}$ &$ 4.72\pm{  0.53}$ &$51.51_{- 0.10}^{+ 0.21}$\\
GRB130610A*&  2.09&   21.76&$ 152.37\pm{  38.74}$ &$-1.10\pm{ 0.29}$ &$ 3.08\pm{  0.33}$ &$52.06_{- 0.33}^{+ 0.23}$\\
GRB130612A*&  2.01&    7.42&$  64.28\pm{  14.52}$ &$-1.20\pm{ 0.39}$ &$ 2.53\pm{  0.26}$ &$51.76_{- 0.19}^{+ 0.29}$\\
GRB131011A*&  1.87&   77.06&$ 402.04\pm{  97.23}$ &$-0.86\pm{ 0.16}$ &$ 4.11\pm{  0.27}$ &$52.38_{- 0.54}^{+ 0.31}$\\
GRB140206A*&  2.73&  146.69&$ 417.65\pm{  13.16}$ &$-0.68\pm{ 0.02}$ &$42.70\pm{  0.43}$ &$53.86_{- 0.05}^{+ 0.05}$\\
GRB140423A*&  3.26&   95.23&$ 250.39\pm{  65.21}$ &$-0.69\pm{ 0.26}$ &$ 2.58\pm{  0.27}$ &$52.65_{- 0.52}^{+ 0.36}$\\
GRB140620A*&  2.04&   45.83&$  92.50\pm{  11.96}$ &$-0.91\pm{ 0.21}$ &$ 6.31\pm{  0.38}$ &$52.23_{- 0.14}^{+ 0.11}$\\
GRB140623A*&  1.92&  111.10&$ 2361.83\pm{ 2361.83}$ &$-1.47\pm{ 0.14}$ &$ 1.49\pm{  0.23}$ &$52.07_{- 0.66}^{+ 0.54}$\\
GRB140703A*&  3.14&   83.97&$ 271.85\pm{  46.25}$ &$-0.77\pm{ 0.15}$ &$ 3.92\pm{  0.23}$ &$52.80_{- 0.26}^{+ 0.20}$\\
GRB140907A*&  1.21&   35.84&$ 113.07\pm{  13.67}$ &$-0.64\pm{ 0.22}$ &$ 3.80\pm{  0.24}$ &$51.51_{- 0.20}^{+ 0.16}$\\
GRB141004A&  0.57&    2.56&$ 140.37\pm{  25.42}$ &$-1.33\pm{ 0.12}$ &$ 8.71\pm{  0.40}$ &$51.08_{- 0.18}^{+ 0.12}$\\
GRB141028A*&  2.33&   31.49&$ 511.30\pm{  34.13}$ &$-0.71\pm{ 0.04}$ &$16.89\pm{  0.30}$ &$53.35_{- 0.12}^{+ 0.09}$\\
GRB141221A*&  1.45&   23.81&$ 328.13\pm{  80.64}$ &$-1.08\pm{ 0.13}$ &$ 5.10\pm{  0.35}$ &$52.08_{- 0.45}^{+ 0.24}$\\
GRB141225A*&  0.92&   56.32&$ 252.79\pm{  46.01}$ &$-0.01\pm{ 0.36}$ &$ 2.21\pm{  0.25}$ &$51.32_{- 0.47}^{+ 0.27}$\\
GRB150120A*&  0.46&    3.33&$ 230.32\pm{ 230.32}$ &$-1.46\pm{ 0.64}$ &$ 4.49\pm{  1.19}$ &$50.64_{- 0.45}^{+ 0.80}$\\
GRB150301B*&  1.52&   13.31&$ 172.35\pm{  26.60}$ &$-0.95\pm{ 0.15}$ &$ 4.44\pm{  0.27}$ &$51.91_{- 0.23}^{+ 0.18}$\\
GRB150727A*&  0.31&   49.41&$ 263.49\pm{  79.74}$ &$-0.53\pm{ 0.46}$ &$ 1.72\pm{  0.30}$ &$50.02_{- 0.62}^{+ 0.41}$\\
GRB160804A*&  0.74&  131.59&$ 133.34\pm{  43.24}$ &$-1.23\pm{ 0.24}$ &$ 3.93\pm{  0.32}$ &$50.99_{- 0.29}^{+ 0.24}$\\
GRB161129A*&  0.65&   36.10&$ 272.60\pm{  56.97}$ &$-0.87\pm{ 0.12}$ &$ 5.21\pm{  0.30}$ &$51.20_{- 0.33}^{+ 0.26}$\\
GRB170214A*&  2.53&  122.88&$ 614.85\pm{  47.48}$ &$-0.78\pm{ 0.04}$ &$17.59\pm{  0.34}$ &$53.51_{- 0.12}^{+ 0.11}$\\
GRB170607A*&  0.56&   20.93&$ 221.14\pm{  36.87}$ &$-1.37\pm{ 0.08}$ &$15.33\pm{  0.51}$ &$51.38_{- 0.16}^{+ 0.14}$\\
GRB170705A*&  2.01&   22.78&$ 207.84\pm{  12.53}$ &$-0.84\pm{ 0.06}$ &$21.73\pm{  0.50}$ &$52.97_{- 0.07}^{+ 0.07}$\\
GRB170817A&  0.01&    2.05&$ 229.17\pm{  78.08}$ &$0.85\pm{ 1.38}$ &$ 3.02\pm{  0.79}$ &$47.21_{- 0.83}^{+ 0.43}$\\
GRB170903A*&  0.89&   25.60&$ 181.58\pm{  52.67}$ &$-1.29\pm{ 0.15}$ &$ 5.44\pm{  0.36}$ &$51.40_{- 0.29}^{+ 0.22}$\\
GRB171222A*&  2.41&   80.38&$ 388.20\pm{   3.60}$ &$-2.02\pm{ 0.01}$ &$ 1.53\pm{  0.25}$ &$52.04_{- 0.03}^{+ 0.49}$\\
GRB180205A*&  1.41&   15.36&$ 493.84\pm{  81.91}$ &$-1.76\pm{ 0.11}$ &$ 3.48\pm{  0.30}$ &$51.82_{- 0.10}^{+ 0.15}$\\
GRB180620B*&  1.12&   46.72&$ 471.61\pm{ 142.94}$ &$-1.20\pm{ 0.13}$ &$ 6.77\pm{  0.51}$ &$51.99_{- 0.56}^{+ 0.32}$\\
GRB180720B*&  0.65&   48.90&$ 928.79\pm{  40.75}$ &$-1.00\pm{ 0.01}$ &$109.80\pm{  0.65}$ &$52.91_{- 0.07}^{+ 0.06}$\\
GRB190114C&  0.42&  170.94&$ 652.37\pm{   3.80}$ &$-1.06\pm{ 0.00}$ &$246.86\pm{  0.86}$ &$52.66_{- 0.01}^{+ 0.01}$\\
  \hline
\end{longtable}
}

{\scriptsize {
 \centering
\begin{longtable}{|c|c|c|c|}
\caption{ 152 Konus/Wind GRBs }
\label{tabKW}\\
  \hline
  GRB   & z & $E_p^{obs}$ & Log($\frac{L_{p}}{\mathrm{erg/s}}$)   \\
  \hline
\endfirsthead
\multicolumn{4}{c}%
{{\bfseries \tablename\ \thetable{} -- continued from previous
page}} \\ \hline
  GRB & z & $E_p^{obs}$ & Log($\frac{L_{p}}{\mathrm{erg/s}}$)     \\
 \hline
\endhead
\hline \multicolumn{4}{|c|}{{Continued on next page}} \\ \hline
\endfoot
\hline \hline
\endlastfoot
   91018& 0.971&$  27.9_{- 13.19}^{+ 13.19}$& $51.13_{- 0.39}^{+ 0.25}$\\
   970228& 0.695&$ 165.2_{- 24.78}^{+ 38.94}$& $52.27_{- 0.14}^{+ 0.18}$\\
   970828& 0.958&$ 271.2_{- 21.96}^{+ 24.01}$& $52.71_{- 0.11}^{+ 0.11}$\\
   971214& 3.418&$ 179.0_{- 16.07}^{+ 19.92}$& $52.99_{- 0.13}^{+ 0.14}$\\
   990123& 1.600&$ 724.1_{- 31.15}^{+ 33.84}$& $53.59_{- 0.13}^{+ 0.13}$\\
   990506& 1.307&$ 296.1_{- 25.15}^{+ 26.88}$& $52.88_{- 0.12}^{+ 0.12}$\\
   990510& 1.619&$ 135.9_{- 9.93}^{+ 11.84}$& $52.43_{- 0.09}^{+ 0.10}$\\
   990705& 0.842&$ 299.1_{- 9.23}^{+ 9.23}$& $52.74_{- 0.07}^{+ 0.07}$\\
   990712& 0.433&$ 104.0_{- 11.16}^{+ 14.65}$& $51.83_{- 0.08}^{+ 0.10}$\\
   991208& 0.706&$ 185.0_{- 4.98}^{+ 4.98}$& $52.35_{- 0.05}^{+ 0.05}$\\
   991216& 1.020&$ 353.0_{- 14.85}^{+ 14.85}$& $52.91_{- 0.09}^{+ 0.09}$\\
   000131& 4.500&$ 133.1_{- 12.00}^{+ 13.09}$& $52.93_{- 0.13}^{+ 0.13}$\\
   000210& 0.846&$ 372.1_{- 20.04}^{+ 21.12}$& $52.89_{- 0.10}^{+ 0.10}$\\
  000301C& 2.034&$ 162.8_{- 35.93}^{+ 65.93}$& $52.66_{- 0.21}^{+ 0.30}$\\
   000418& 1.118&$ 116.1_{- 8.03}^{+ 9.91}$& $52.18_{- 0.07}^{+ 0.08}$\\
   000911& 1.058&$ 1082.8_{- 50.04}^{+ 51.98}$& $53.71_{- 0.13}^{+ 0.14}$\\
  000926A& 2.037&$ 108.0_{- 6.91}^{+ 7.90}$& $52.38_{- 0.08}^{+ 0.09}$\\
   010222& 1.477&$ 285.0_{- 20.19}^{+ 21.80}$& $52.91_{- 0.11}^{+ 0.12}$\\
   010921& 0.450&$  93.1_{- 6.90}^{+ 8.28}$& $51.76_{- 0.05}^{+ 0.06}$\\
   011121& 0.360&$ 819.1_{- 96.32}^{+ 108.09}$& $53.22_{- 0.16}^{+ 0.17}$\\
   020405& 0.690&$ 161.0_{- 7.10}^{+ 7.10}$& $52.24_{- 0.06}^{+ 0.06}$\\
   020813& 1.254&$ 227.2_{- 16.86}^{+ 18.19}$& $52.68_{- 0.10}^{+ 0.11}$\\
  020819B& 0.411&$ 158.8_{- 28.35}^{+ 41.81}$& $52.11_{- 0.15}^{+ 0.19}$\\
   030329& 0.169&$  97.0_{- 1.97}^{+ 1.97}$& $51.64_{- 0.01}^{+ 0.01}$\\
   031203& 0.105&$ 143.0_{- 46.15}^{+ 46.15}$& $51.87_{- 0.27}^{+ 0.21}$\\
   041006& 0.716&$  86.0_{- 5.01}^{+ 5.01}$& $51.82_{- 0.05}^{+ 0.05}$\\
   050401& 2.899&$ 104.9_{- 11.03}^{+ 14.11}$& $52.53_{- 0.12}^{+ 0.13}$\\
  050525A& 0.606&$  80.0_{- 1.99}^{+ 1.99}$& $51.72_{- 0.02}^{+ 0.02}$\\
   050603& 2.821&$ 238.9_{- 34.02}^{+ 64.12}$& $53.09_{- 0.17}^{+ 0.24}$\\
   050820& 2.615&$ 485.0_{- 92.12}^{+ 149.11}$& $53.54_{- 0.23}^{+ 0.29}$\\
   051008& 2.770&$ 550.1_{- 93.90}^{+ 132.89}$& $53.66_{- 0.22}^{+ 0.26}$\\
   051022& 0.800&$ 441.1_{- 16.11}^{+ 17.78}$& $52.99_{- 0.09}^{+ 0.09}$\\
  051109A& 2.346&$ 170.1_{- 42.14}^{+ 80.10}$& $52.76_{- 0.24}^{+ 0.34}$\\
   060124& 2.297&$ 239.0_{- 26.08}^{+ 33.97}$& $52.98_{- 0.14}^{+ 0.16}$\\
   060418& 1.489&$ 229.8_{- 57.45}^{+ 57.45}$& $52.76_{- 0.24}^{+ 0.22}$\\
  060502A& 1.503&$ 127.9_{- 25.17}^{+ 45.95}$& $52.36_{- 0.18}^{+ 0.26}$\\
   060814& 1.923&$ 430.1_{- 126.93}^{+ 360.94}$& $53.31_{- 0.31}^{+ 0.53}$\\
  060912A& 0.937&$ 203.9_{- 74.86}^{+ 304.08}$& $52.50_{- 0.34}^{+ 0.71}$\\
   061007& 1.261&$ 398.9_{- 11.06}^{+ 11.94}$& $53.08_{- 0.09}^{+ 0.09}$\\
   061021& 0.345&$ 521.8_{- 139.00}^{+ 265.37}$& $52.90_{- 0.27}^{+ 0.36}$\\
   061121& 1.314&$ 607.2_{- 45.81}^{+ 51.86}$& $53.39_{- 0.14}^{+ 0.15}$\\
  061222A& 2.088&$ 297.9_{- 22.02}^{+ 25.91}$& $53.09_{- 0.12}^{+ 0.13}$\\
   070125& 1.547&$ 371.8_{- 31.02}^{+ 36.12}$& $53.11_{- 0.13}^{+ 0.14}$\\
   070328& 2.063&$ 385.9_{- 47.02}^{+ 59.10}$& $53.27_{- 0.17}^{+ 0.18}$\\
   070508& 0.820&$ 188.0_{- 4.01}^{+ 5.00}$& $52.40_{- 0.05}^{+ 0.05}$\\
   070521& 2.087&$ 190.8_{- 12.31}^{+ 13.93}$& $52.78_{- 0.10}^{+ 0.11}$\\
   071003& 1.604&$ 801.0_{- 62.97}^{+ 72.19}$& $53.66_{- 0.15}^{+ 0.16}$\\
  071010B& 0.947&$  55.0_{- 4.98}^{+ 4.01}$& $51.60_{- 0.06}^{+ 0.05}$\\
   071020& 2.146&$ 322.0_{- 34.01}^{+ 43.86}$& $53.16_{- 0.15}^{+ 0.17}$\\
  071112C& 0.823&$ 406.0_{- 99.85}^{+ 178.86}$& $52.94_{- 0.25}^{+ 0.33}$\\
   071117& 1.331&$ 278.0_{- 54.92}^{+ 100.82}$& $52.85_{- 0.20}^{+ 0.28}$\\
  080319B& 0.938&$ 652.2_{- 8.77}^{+ 8.77}$& $53.31_{- 0.09}^{+ 0.09}$\\
  080319C& 1.949&$ 632.0_{- 112.91}^{+ 160.04}$& $53.58_{- 0.23}^{+ 0.26}$\\
   080411& 1.030&$ 266.0_{- 18.23}^{+ 21.18}$& $52.72_{- 0.10}^{+ 0.11}$\\
  080413A& 2.433&$ 192.0_{- 43.11}^{+ 110.98}$& $52.86_{- 0.23}^{+ 0.39}$\\
  080514B& 1.800&$ 196.1_{- 12.14}^{+ 12.14}$& $52.73_{- 0.09}^{+ 0.10}$\\
   080602& 1.820&$ 399.9_{- 128.00}^{+ 738.90}$& $53.23_{- 0.33}^{+ 0.84}$\\
  080603B& 2.689&$ 101.1_{- 17.89}^{+ 37.14}$& $52.46_{- 0.17}^{+ 0.27}$\\
   080605& 1.640&$ 259.8_{- 9.09}^{+ 9.85}$& $52.89_{- 0.08}^{+ 0.09}$\\
   080607& 3.036&$ 334.0_{- 16.10}^{+ 17.09}$& $53.36_{- 0.12}^{+ 0.12}$\\
   080721& 2.591&$ 490.1_{- 40.94}^{+ 40.94}$& $53.54_{- 0.15}^{+ 0.15}$\\
   080913& 6.695&$  92.3_{- 45.48}^{+ 45.48}$& $52.91_{- 0.51}^{+ 0.35}$\\
  080916A& 0.689&$ 129.1_{- 14.21}^{+ 17.77}$& $52.09_{- 0.10}^{+ 0.11}$\\
  080916C& 4.350&$ 505.0_{- 42.99}^{+ 45.98}$& $53.84_{- 0.17}^{+ 0.17}$\\
   081121& 2.512&$ 254.0_{- 17.94}^{+ 21.92}$& $53.07_{- 0.12}^{+ 0.13}$\\
  081203A& 2.050&$ 438.0_{- 162.95}^{+ 607.87}$& $53.35_{- 0.39}^{+ 0.72}$\\
   081221& 2.260&$  81.0_{- 1.99}^{+ 1.99}$& $52.22_{- 0.04}^{+ 0.04}$\\
   081222& 2.770&$ 192.0_{- 19.10}^{+ 24.93}$& $52.93_{- 0.13}^{+ 0.15}$\\
   090102& 1.547&$ 431.9_{- 36.12}^{+ 42.01}$& $53.22_{- 0.13}^{+ 0.15}$\\
   090201& 2.100&$ 156.1_{- 6.13}^{+ 7.10}$& $52.65_{- 0.07}^{+ 0.08}$\\
   090323& 3.570&$ 417.1_{- 45.08}^{+ 45.08}$& $53.60_{- 0.17}^{+ 0.17}$\\
   090328& 0.736&$ 747.1_{- 86.98}^{+ 107.72}$& $53.33_{- 0.16}^{+ 0.18}$\\
   090424& 0.544&$ 162.0_{- 6.02}^{+ 6.02}$& $52.19_{- 0.05}^{+ 0.05}$\\
   090618& 0.540&$ 187.0_{- 5.97}^{+ 5.97}$& $52.28_{- 0.05}^{+ 0.05}$\\
  090709A& 1.800&$ 277.1_{- 10.00}^{+ 10.00}$& $52.97_{- 0.09}^{+ 0.09}$\\
  090715B& 3.000&$ 135.0_{- 19.00}^{+ 26.00}$& $52.72_{- 0.15}^{+ 0.18}$\\
   090812& 2.452&$ 384.1_{- 104.87}^{+ 263.04}$& $53.35_{- 0.29}^{+ 0.46}$\\
  090926A& 2.106&$ 327.1_{- 8.05}^{+ 8.05}$& $53.16_{- 0.09}^{+ 0.09}$\\
  100606A& 1.554&$ 874.1_{- 160.89}^{+ 241.93}$& $53.71_{- 0.24}^{+ 0.28}$\\
  100621A& 0.542&$ 105.7_{- 7.78}^{+ 9.08}$& $51.89_{- 0.06}^{+ 0.07}$\\
  100728A& 1.567&$ 305.0_{- 15.97}^{+ 15.97}$& $52.98_{- 0.10}^{+ 0.10}$\\
  100814A& 1.440&$ 127.9_{- 11.07}^{+ 13.11}$& $52.34_{- 0.09}^{+ 0.10}$\\
  101213A& 0.414&$ 250.4_{- 29.70}^{+ 41.02}$& $52.43_{- 0.12}^{+ 0.15}$\\
  110213B& 1.083&$ 122.9_{- 19.20}^{+ 19.20}$& $52.20_{- 0.14}^{+ 0.13}$\\
  110422A& 1.770&$ 155.0_{- 3.00}^{+ 3.00}$& $52.56_{- 0.06}^{+ 0.06}$\\
  110503A& 1.613&$ 220.1_{- 11.86}^{+ 11.86}$& $52.77_{- 0.09}^{+ 0.09}$\\
  110715A& 0.820&$ 119.2_{- 7.14}^{+ 7.14}$& $52.09_{- 0.06}^{+ 0.06}$\\
  110731A& 2.830&$ 288.0_{- 15.93}^{+ 18.02}$& $53.22_{- 0.11}^{+ 0.12}$\\
  110918A& 0.984&$ 336.2_{- 34.78}^{+ 39.82}$& $52.87_{- 0.13}^{+ 0.14}$\\
  111008A& 5.000&$ 104.0_{- 20.00}^{+ 31.00}$& $52.82_{- 0.20}^{+ 0.25}$\\
  111228A& 0.716&$  43.1_{- 30.89}^{+ 18.07}$& $51.34_{- 0.81}^{+ 0.24}$\\
  120119A& 1.728&$ 152.9_{- 12.10}^{+ 12.10}$& $52.54_{- 0.10}^{+ 0.10}$\\
  120624B& 2.197&$ 560.1_{- 41.91}^{+ 49.10}$& $53.55_{- 0.15}^{+ 0.16}$\\
  120711A& 1.405&$ 1061.1_{- 36.17}^{+ 37.84}$& $53.80_{- 0.13}^{+ 0.13}$\\
  120716A& 2.486&$ 189.9_{- 30.12}^{+ 49.91}$& $52.86_{- 0.17}^{+ 0.23}$\\
  120909A& 3.930&$ 335.1_{- 24.95}^{+ 24.95}$& $53.50_{- 0.14}^{+ 0.14}$\\
  121128A& 2.200&$  76.9_{- 3.12}^{+ 3.12}$& $52.18_{- 0.05}^{+ 0.05}$\\
  130408A& 3.758&$ 270.9_{- 43.09}^{+ 36.99}$& $53.33_{- 0.20}^{+ 0.18}$\\
  130427A& 0.340&$ 1056.0_{- 9.70}^{+ 9.70}$& $53.39_{- 0.09}^{+ 0.09}$\\
  130505A& 2.270&$ 593.0_{- 23.85}^{+ 25.99}$& $53.61_{- 0.12}^{+ 0.13}$\\
  130514A& 3.600&$ 110.0_{- 21.09}^{+ 41.96}$& $52.68_{- 0.19}^{+ 0.29}$\\
  130518A& 2.488&$ 332.0_{- 45.01}^{+ 53.90}$& $53.25_{- 0.18}^{+ 0.19}$\\
  130606A& 5.913&$ 293.9_{- 50.05}^{+ 89.98}$& $53.64_{- 0.22}^{+ 0.29}$\\
  130701A& 1.155&$  89.0_{- 3.99}^{+ 3.99}$& $52.00_{- 0.05}^{+ 0.05}$\\
  130831A& 0.479&$  54.0_{- 8.99}^{+ 7.03}$& $51.39_{- 0.10}^{+ 0.07}$\\
  130907A& 1.238&$ 387.0_{- 16.09}^{+ 16.09}$& $53.05_{- 0.10}^{+ 0.10}$\\
  130925A& 0.347&$ 181.1_{- 9.65}^{+ 9.65}$& $52.17_{- 0.06}^{+ 0.06}$\\
  131030A& 1.293&$ 195.8_{- 6.11}^{+ 6.11}$& $52.59_{- 0.07}^{+ 0.07}$\\
  131105A& 1.686&$ 201.0_{- 26.06}^{+ 39.84}$& $52.72_{- 0.14}^{+ 0.18}$\\
  131108A& 2.400&$ 357.9_{- 25.88}^{+ 30.88}$& $53.29_{- 0.13}^{+ 0.14}$\\
  131231A& 0.644&$ 161.8_{- 6.08}^{+ 6.08}$& $52.23_{- 0.05}^{+ 0.05}$\\
  140213A& 1.208&$ 100.0_{- 3.99}^{+ 3.99}$& $52.10_{- 0.05}^{+ 0.05}$\\
 140226A*& 1.980&$ 414.1_{- 78.86}^{+ 78.86}$& $53.30_{- 0.22}^{+ 0.21}$\\
  140419A& 3.956&$ 436.0_{- 39.95}^{+ 49.03}$& $53.69_{- 0.16}^{+ 0.18}$\\
  140506A& 0.889&$ 200.1_{- 41.82}^{+ 89.99}$& $52.47_{- 0.19}^{+ 0.31}$\\
  140508A& 1.027&$ 220.0_{- 12.83}^{+ 13.81}$& $52.59_{- 0.08}^{+ 0.09}$\\
  140512A& 0.725&$ 478.8_{- 78.84}^{+ 117.10}$& $53.02_{- 0.19}^{+ 0.23}$\\
  140606B& 0.384&$ 254.3_{- 26.73}^{+ 33.24}$& $52.42_{- 0.11}^{+ 0.13}$\\
  140629A& 2.275&$  86.1_{- 17.10}^{+ 17.10}$& $52.27_{- 0.17}^{+ 0.16}$\\
  140801A& 1.320&$ 108.0_{- 3.02}^{+ 3.02}$& $52.19_{- 0.04}^{+ 0.04}$\\
  140808A& 3.293&$ 125.1_{- 10.95}^{+ 13.98}$& $52.72_{- 0.11}^{+ 0.13}$\\
  141109A& 2.993&$ 191.1_{- 43.08}^{+ 75.88}$& $52.96_{- 0.23}^{+ 0.31}$\\
  141220A& 1.319&$ 138.8_{- 9.05}^{+ 9.92}$& $52.36_{- 0.08}^{+ 0.08}$\\
  150206A& 2.087&$ 228.1_{- 22.03}^{+ 23.00}$& $52.91_{- 0.13}^{+ 0.13}$\\
  150314A& 1.758&$ 349.9_{- 10.15}^{+ 10.15}$& $53.12_{- 0.09}^{+ 0.09}$\\
  150323A& 0.593&$  94.8_{- 8.16}^{+ 8.79}$& $51.84_{- 0.07}^{+ 0.07}$\\
  150403A& 2.060&$ 372.9_{- 31.05}^{+ 33.99}$& $53.24_{- 0.14}^{+ 0.14}$\\
  150413A& 3.139&$  95.9_{- 21.99}^{+ 56.05}$& $52.51_{- 0.21}^{+ 0.38}$\\
  150818A& 0.282&$  99.8_{- 17.94}^{+ 28.86}$& $51.72_{- 0.13}^{+ 0.19}$\\
  150821A& 0.755&$ 435.9_{- 71.79}^{+ 107.12}$& $52.96_{- 0.18}^{+ 0.22}$\\
  151021A& 2.330&$ 170.0_{- 11.11}^{+ 12.91}$& $52.75_{- 0.10}^{+ 0.11}$\\
  151027A& 0.810&$ 172.9_{- 32.04}^{+ 58.01}$& $52.34_{- 0.17}^{+ 0.24}$\\
  160131A& 0.972&$ 651.1_{- 156.19}^{+ 230.22}$& $53.32_{- 0.26}^{+ 0.30}$\\
  160509A& 1.170&$ 288.0_{- 27.19}^{+ 29.03}$& $52.82_{- 0.12}^{+ 0.13}$\\
  160623A& 0.367&$ 552.3_{- 13.90}^{+ 13.90}$& $52.95_{- 0.08}^{+ 0.08}$\\
  160625B& 1.406&$ 571.1_{- 10.81}^{+ 12.05}$& $53.37_{- 0.10}^{+ 0.10}$\\
  160629A& 3.332&$ 235.9_{- 18.93}^{+ 21.93}$& $53.16_{- 0.13}^{+ 0.14}$\\
  161014A& 2.823&$ 226.0_{- 79.00}^{+ 267.07}$& $53.05_{- 0.35}^{+ 0.64}$\\
  161017A& 2.013&$ 289.1_{- 73.02}^{+ 73.02}$& $53.05_{- 0.26}^{+ 0.23}$\\
  161023A& 2.708&$ 162.9_{- 28.05}^{+ 36.95}$& $52.80_{- 0.18}^{+ 0.21}$\\
  161117A& 1.549&$  69.0_{- 7.06}^{+ 5.88}$& $51.94_{- 0.08}^{+ 0.07}$\\
  170202A& 3.645&$ 246.9_{- 85.90}^{+ 165.98}$& $53.24_{- 0.36}^{+ 0.45}$\\
  170604A& 1.329&$ 219.8_{- 48.09}^{+ 72.13}$& $52.68_{- 0.21}^{+ 0.26}$\\
  171010A& 0.329&$ 171.0_{- 6.02}^{+ 7.00}$& $52.12_{- 0.04}^{+ 0.05}$\\
  171205A& 0.037&$ 120.6_{- 35.69}^{+ 136.00}$& $51.70_{- 0.23}^{+ 0.55}$\\
  180325A& 2.248&$ 306.0_{- 39.10}^{+ 49.88}$& $53.15_{- 0.16}^{+ 0.18}$\\
  180728A& 0.117&$  96.7_{- 6.27}^{+ 7.16}$& $51.60_{- 0.04}^{+ 0.05}$\\
 180914B*& 1.096&$ 466.1_{- 27.19}^{+ 29.10}$& $53.13_{- 0.11}^{+ 0.12}$\\
  181020A& 2.938&$ 371.0_{- 52.06}^{+ 57.14}$& $53.41_{- 0.19}^{+ 0.19}$\\
  181110A& 1.505&$  47.9_{- 27.15}^{+ 13.97}$& $51.68_{- 0.55}^{+ 0.19}$\\
  181201A& 0.450&$ 151.7_{- 6.21}^{+ 6.21}$& $52.10_{- 0.05}^{+ 0.05}$\\
\hline
\end{longtable}
} }
\end{document}